\documentclass[prx,superscriptaddress,amsfonts,amssymb,amsmath,floats,twocolumn,aps,footinbib,shownopacs]{revtex4}
\usepackage[pdftex]{graphicx}
\usepackage{subfigure}
\usepackage{dsfont}
\usepackage{lipsum}
\usepackage{dcolumn}
\usepackage{bm}
\usepackage{color}
\usepackage{physics}
\usepackage{multirow}
\usepackage[pdftex,colorlinks=true, linkcolor=blue,citecolor=blue,filecolor=blue]{hyperref}
\begin{document}
\title{Inhomogeneous disordering at a photo-induced charge density wave transition }

\author{Antonio Picano} 
\affiliation{Department of Physics, University of Erlangen-Nuremberg, Staudtstra{\ss}e 7, 91058 Erlangen, Germany}
\author{Francesco Grandi} 
\affiliation{Department of Physics, University of Erlangen-Nuremberg, Staudtstra{\ss}e 7, 91058 Erlangen, Germany}
\author{Martin Eckstein}
\affiliation{Department of Physics, University of Erlangen-Nuremberg, Staudtstra{\ss}e 7, 91058 Erlangen, Germany}

\begin{abstract}
Using ultrashort laser pulses, it has become possible to probe the dynamics of long-range order in solids on microscopic timescales. In the conventional description of symmetry-broken phases within time-dependent Ginzburg-Landau theory, the order parameter evolves coherently, with small fluctuations along an average trajectory. Recent experiments, however, indicate that some systems can support a different scenario, named ultrafast inhomogeneous disordering, where the average order parameter is no longer representative of the state on the atomic scale. Here we theoretically show that ultrafast disordering can occur in a minimal, yet paradigmatic, model for a Peierls instability if atomic scale inhomogeneities of both the electronic structure and the charge density  wave order parameter are taken into account. The latter is achieved using a non-equilibrium generalization of statistical dynamical mean-field theory, coupled to stochastic differential equations for the order parameter.
\end{abstract} 
\maketitle

\section{Introduction}

Symmetry breaking phase transitions are among the most fundamental phenomena in physics, from cosmology to condensed matter. Understanding their dynamics in solids is therefore of basic interest as much as it is needed to establish pathways to control complex states of matter on ultrafast timescales \cite{Basov2017, delaTorre2021}. The conventional phenomenological understanding of symmetry breaking is based on Ginzburg-Landau theory (GLT), which determines the order parameter from a free energy density. The latter depends on the electronic state through few variables like temperature, and it can therefore be rapidly modified by an excitation of the electrons. Time-dependent GLT has been successfully used to describe the resulting coherent dynamics of various  orders, including superconductivity and charge density waves \cite{Yusupov2010, Schaefer2010, Trigo2019, Huber2014, Neugebauer2019, Beaud2014, Zong2019}. Nevertheless, in many experiments the dynamics of ordered phases comes with yet unresolved mysteries, including the emergence of metastable states which cannot be reached along equilibrium pathways \cite{Ichikawa2011, Stojchevska2014, Budden2021}, or dynamics which is not governed by the free energy for the measured electronic temperature \cite{Maklar2021}. 

A theoretical description beyond time-dependent GLT must properly include non-thermal order parameter fluctuations on various length and time scales. Including spatial fluctuations into time-dependent GLT in fact has a profound influence on the dynamics. For example, fluctuations of a thermodynamically subdominant order can become observable when the dominant order is transiently suppressed \cite{Zong2021}, and they can be important for a transition to metastable states \cite{Sun2020}. Moreover, due to the anharmonicity of the potential, non-thermal fluctuations can renormalize the free-energy. This can lead to a slow-down of the dynamics \cite{Dolgirev2020}, or, in systems with discrete symmetry breaking, to a qualitative change of the potential \cite{Grandi2021}. Finally, even in the disordered phase, non-thermal order parameter fluctuations can leave characteristic signatures in the electronic properties  \cite{Bauer2015, Lemonik2017, Lemonik2018b, Stahl2021} 

\begin{figure*}
\centerline{\includegraphics[width=1\textwidth]{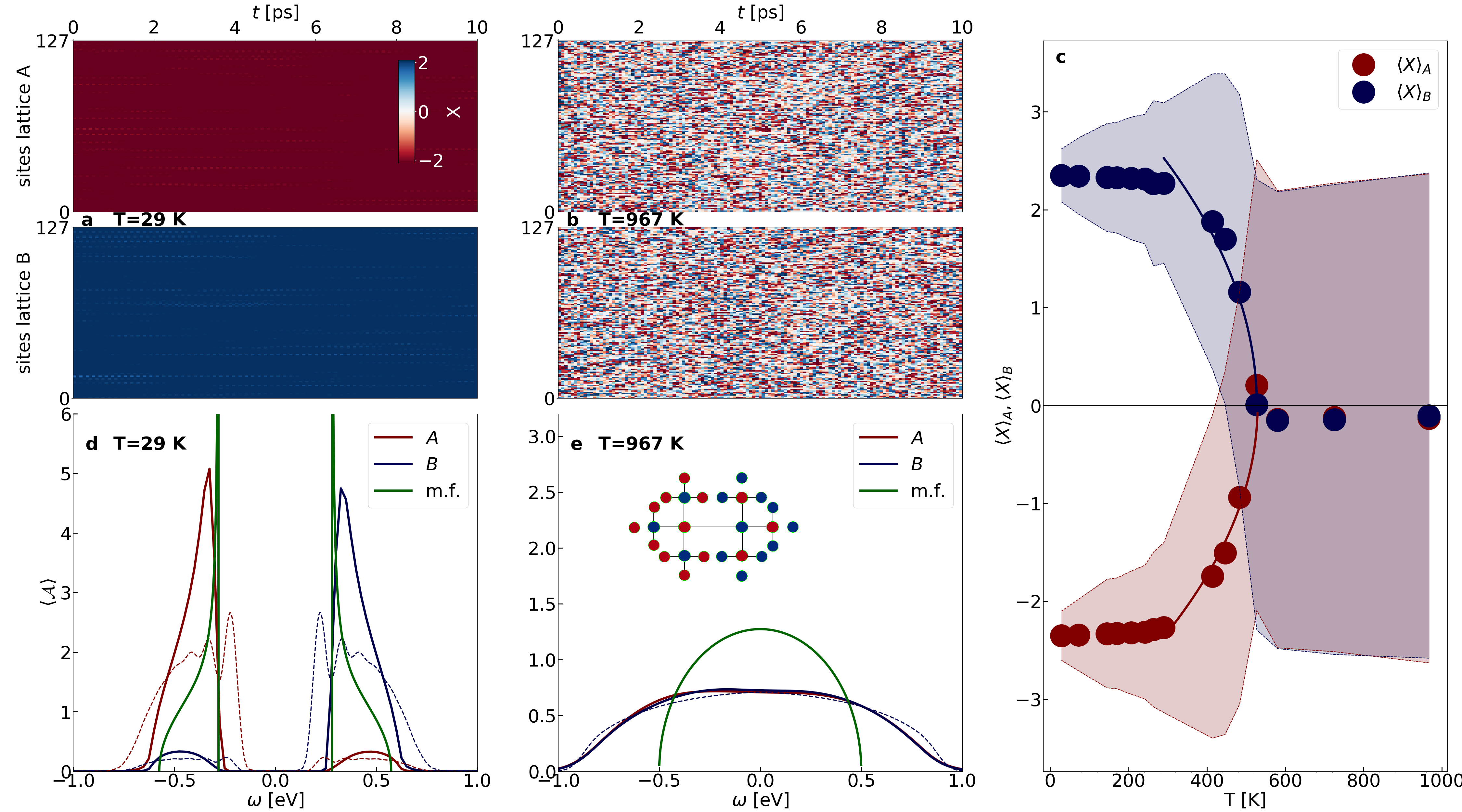}}
\caption{{\bf Equilibrium solution of the model} --- \textbf{a}) and \textbf{b})  Stochastic evolution of $X_j$ in the stationary equilibrium phase at a temperature $T = 29$K corresponding to the ordered phase ({\bf a}) and at $T = 967$K in the disordered phase ({\bf b}). The vertical axis labels the $256$ representative sites $j$, grouping together sites of the same sublattice. Note that there is no correlation between the representative sites, their labelling is arbitrary. \textbf{c}) Average $\langle X \rangle_\alpha$ for the two sublattices $\alpha = A, B$ as a function of temperature (dots). Solid lines correspond to a fit $\langle X \rangle_{A,B} = \pm C \sqrt{T_c  - T}$ in the ordered phase, with $T_c \approx 527$K. The shaded areas represent the confidence interval of each $\langle X \rangle_\alpha$, with semi-amplitude given by the variance $\sigma_\alpha = \sqrt{\langle X^2 \rangle_\alpha - \langle X \rangle_\alpha^2}$. $\langle X \rangle_\alpha$ has to be intended as an average over all the $N=128$ sites in sublattice $\alpha=A,B$, and over all the times: $\langle X \rangle_\alpha \equiv E_t [\frac{1}{N} \sum_{j=1}^{N} X_{j,\alpha}(t)]$, where $E_t[\dots]$ is the expectation value over time. \textbf{d}) and \textbf{e}) Electronic spectra at $T = 29$K and $T = 967$K, respectively. Continuous blue and red lines are the average spectral functions taken over all the impurities belonging to a given sublattice. Green lines show the results of a static mean-field solution with the same order parameter as obtained within the stochastic approach and dashed blue and red lines correspond to the perturbative DMFT solution of Ref.~\cite{Randi2017} (see discussion). The inset in \textbf{e} shows a portion of the Bethe lattice with coordination number $4$.}
\label{fig:equi}
\end{figure*}

In the above mentioned extensions of GLT, the order parameter is described by a homogeneous time-dependent mean $\phi_0 (t)$, with small spacial fluctuations $\delta \phi (\vec{r},t)$ that are treated within a Gaussian approximation, i.e. the distribution of the fluctuations is assumed to be Gaussian around the average $\phi_0 (t)$. An entirely different paradigm, which has been put forward in recent experimental studies \cite{Wall2018, Perez-Salinas2022_NatComm, Johnson2022_PRL}, is ultrafast inhomogeneous disordering: In this scenario, the local configuration of the order parameter has a highly non-Gaussian distribution and is therefore no longer represented by its average $\phi_0 (t)$. For example, in a discrete ($\mathbb{Z}_2$) symmetry-breaking transition, ultrafast inhomogeneous disordering could imply that the local order parameter shows a transient bimodal distribution peaked around large positive and negative displacements, say $\phi_0 (t) + \Delta \phi$ and $\phi_0 (t) - \Delta \phi$, respectively. In this case, its average $\phi_0 (t)$ is not representative anymore of the distribution of the single displacements. Although this state cannot be distinguished from a Gaussian disordered state on the macroscopic level, by just looking at the average order parameter,  its local microscopic nature is not captured by the Gaussian approximation, and one can expect a profoundly different dynamics. From a different perspective, this behavior corresponds to a temporary high density of atomic scale defects in the ordered state.

Inhomogeneous disordering also questions the conventional assumption that, after a short relaxation, electrons can be described by a few variables like the effective temperature and the excitation density. Along with the order parameter, in fact, also the local electronic structure may not be well represented by averaged spectra and distribution functions. Moreover, rapid electronic thermalization can be inhibited in disordered systems even in the presence of interactions, as for the case of many-body localization \cite{Nandkishore2015, Abanin2019}. The intertwined evolution of the order parameter fluctuations and the electronic structure may lead to unusually slow relaxation, reminiscent of  weak ergodicity breaking in  translationally invariant systems due to dynamical bottlenecks and constraints \cite{Carleo2012, Smith2017a, Yao2016, Lan2018, Horssen2015}. A theoretical description of inhomogeneous disordering should therefore consider the possibility of a non-thermal and spatially varying electronic state, while previous simulations of dynamical symmetry breaking which start from a microscopic description of the electrons often neglect spacial fluctuations  \cite{Kemper2015, Sentef2016, Werner2012, Tsuji2013}, with a recent exception \cite{Seo2018}.

With the current state of theory, even a minimal model to study inhomogeneous disordering should provide important insights. In the present work, we solve the coupled equations for the inhomogeneous order parameter evolution and the non-equilibrium electron dynamics in a minimal, yet paradigmatic model for a charge ordering transition, i.e., the Holstein model on the infinitely coordinated Bethe lattice. We find that already this simple model supports an inhomogeneous disordering scenario, which suggests that such self-generated disorder might  more generally be of importance for the photo-induced dynamics in solids. Our model also allows to investigate experimentally relevant aspects of that state, in particular the in-gap spectral weight and the slow recovery dynamics.

The article is structured as follows: In Sec.~\ref{sec:model}, we introduce the Holstein model and the approach we developed to take into account non-equilibrium fluctuations of the order parameter. Section \ref{sec:num_sim} presents both equilibrium results and non-equilibrium simulations after a photoexcitation. In Sec.~\ref{sec:noneq_pot}, we reconstruct some out of equilibrium features of the potential energy, i.e., the potential energy barrier, without any original assumption on its existence. Finally, Sec.~\ref{sec:concl} is devoted to concluding remarks.

\begin{figure}
\centering \includegraphics[width=0.5\textwidth]{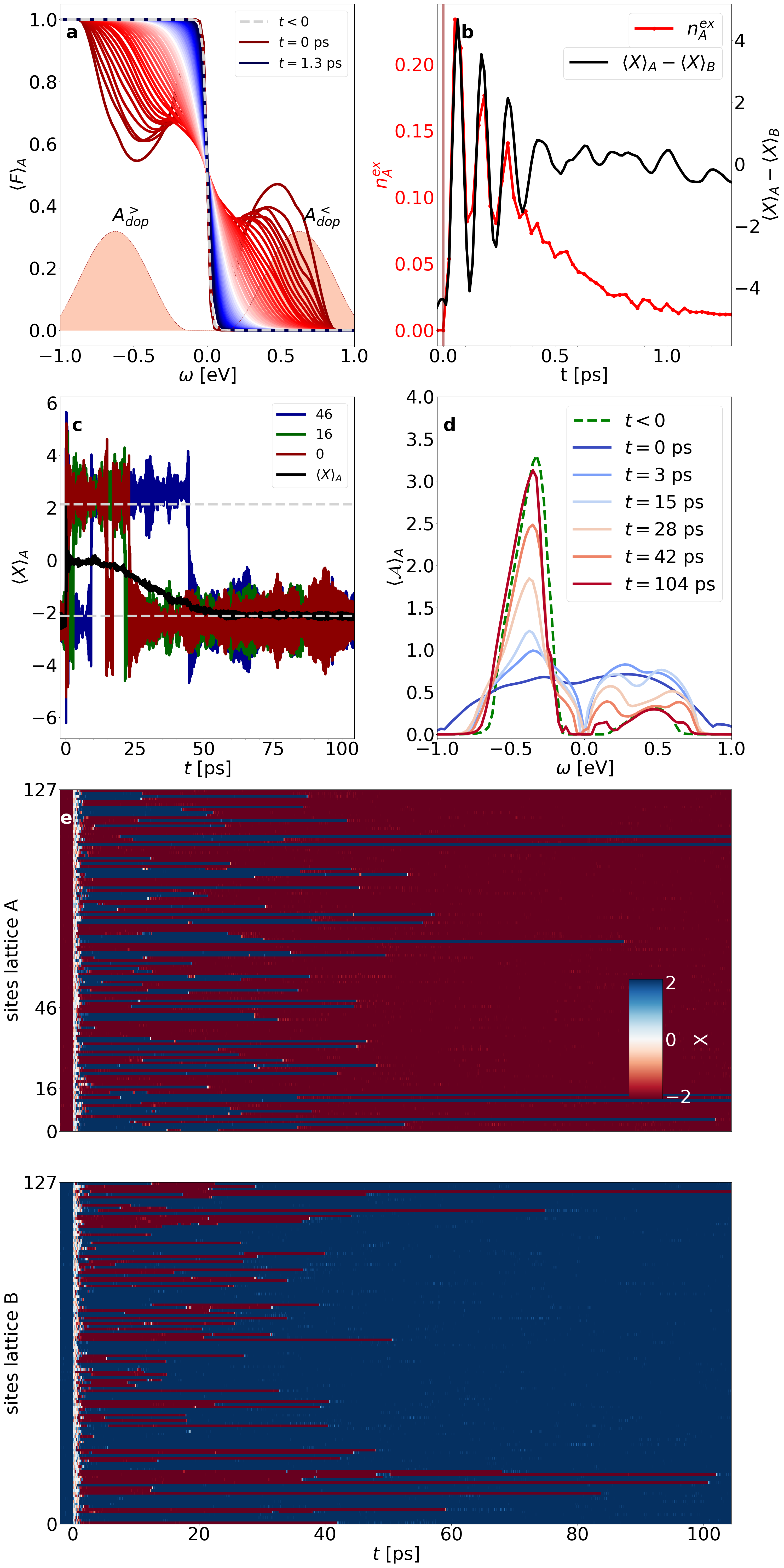}
\caption{{\bf Nonequilibrium evolution of the system} --- \textbf{a}) Time evolution of the electronic distribution function starting from the equilibrium ordered state at $T = 145$K. The excitation is realized by coupling the system for a period of few fs to a photo-doping reservoir with occupied (unoccupied) density of state $\mathcal{A}_{\rm dop}^<$ ($\mathcal{A}_{\rm dop}^>$) above (below) the Fermi energy (shaded areas). \textbf{b}) Short time evolution of the difference between the sublattice averages $\langle X\rangle_A - \langle X\rangle_B$ and of the electronic excitation density $n_A^{\text{ex}} (t)$. \textbf{c}) Long time evolution of the average $\langle X\rangle_A$, together with some representative trajectories $X_j$. \textbf{d}) Time evolution of the average spectral function $\langle\mathcal{A}(\omega, t ) \rangle_{A}$ for the $A$ sublattice. \textbf{e}) Time evolution of $X_j$ after the photoexcitation at $t=0$ for all trajectories.}
\label{fig:noneq}
\end{figure}

\section{Model and theoretical approach} \label{sec:model}
We start from the Holstein Hamiltonian
\begin{align} \label{hamilt}
	H=
	-\!\!
	\sum_{ \langle i,j\rangle ,\sigma} 
	J_{ij}
	c_{i,\sigma}^\dagger c_{j,\sigma}
	+
 	\sum_{j}
	\sqrt{2} g(n_j-1) X_j
	+H_{ph}.
\end{align}
The first term describes tunneling of electrons with hopping amplitude $J_{ij}$ between nearest-neighbor sites $i$ and $j$, and the second term couples the displacement $X_j$ of a local oscillator with the electron density $n_j=n_{j,\uparrow}+n_{j,\downarrow}$; $c_{j,\sigma}$ ($c_{j,\sigma}^\dagger$) are annihilation (creation) operators for electrons with spin $\sigma$ on the lattice site $j$, and $H_{ph}= \sum_{j} \frac{\Omega}{2}(X_j^2+P_j^2)$ is the Hamiltonian of the free oscillators at each site (Einstein phonon). Because we have in mind an evolution over several picoseconds in a solid state environment, we also add a coupling to a thermal reservoir, so that electrons can pass energy to variables other than the particular mode $X$. The bath is included via a dissipative self-energy (see App.~\ref{sec:stat_dmft}, \ref{sec:sto_eq} and \ref{sec:qbe}).

On a bipartite lattice with sub-lattices $A$ and $B$, the model favors a symmetry-broken low-temperature state at half filling, with opposite displacement $\langle X_j\rangle = \pm X_0$ for sites $j$ on the two sublattices, and a gap in the electronic spectrum. We consider the model on an infinitely coordinated Bethe lattice at half filling, which allows for an exact solution within dynamical mean-field theory (DMFT) \cite{Georges1996, Aoki2014}. The noninteracting electronic density of states has a semi-elliptic shape $D(\epsilon) = \frac{4}{\pi W^2}\sqrt{W^2- 4 \epsilon^2}$ with bandwidth $W$. One can analyse the dynamics after an excitation in terms of the local displacements $X_j(t)$ and the local electronic Green's function $G_j(t,t')$, which determines the local spectral function $\mathcal{A}_j (\omega,t) = -\frac{1}{\pi} \text{Im} G^R_j (\omega,t)$, and the local distribution function $F_j ( \omega, t ) = G^<_j (\omega, t)/(2 \pi i\mathcal{A}_j (\omega,t))$. A DMFT solution which enforces a spatially homogeneous order parameter and treats the electron-lattice interaction perturbatively gives coherent order-parameter oscillations after an impulsive electronic excitation, as qualitatively expected from time-dependent GLT \cite{Randi2017}. Here, we allow for an arbitrary distribution of $X_j$ and $G_j$. The non-perturbative solution of the  dynamics in this case is facilitated by the following steps (additional details are reported in the Appendix): (i) Because the relevant phonon timescale $1/\Omega$ is slow compared to the electronic timescale, the exact Keldysh action for the oscillator displacement $X_j$ at a given site $j$ can be replaced by a stochastic equation of motion \cite{KamenevBook}. The coupling of the oscillator to the electronic density fluctuations is replaced by a damping $- \gamma_j (t) \dot{X}_j (t)$ and a stochastic force $\xi_j (t)$ in the white noise limit $\langle \xi_j (t) \rangle = 0$ and $\langle \xi_j (t) \xi_{j'} (t') \rangle = K_j (t) \delta_{j,j'} \delta (t-t')$ (further details about the derivation of these contributions to the equation of motion can be found in App.~\ref{sec:sto_eq} and in Ref.~\cite{Picano2022_arXiv}). The total force on the oscillator at site $j$ is therefore
 \begin{align}
 \label{force}
	f_j = - \Omega^2 X_j - \sqrt{2} g \Omega (\langle n_j\rangle -1 )- \gamma_j \dot X_j + \sqrt{\Omega} \xi_j,
\end{align}
where the first two terms are the Hooke's law and the standard mean-field (Ehrenfest) force. The damping constant $\gamma_j$ and the noise amplitude $K_j$ are determined self-consistently by the retarded and Keldysh components of the local electronic density-density correlation function at site $j$ \cite{KamenevBook}. (ii) After the phonon is replaced by the stochastic variable $X_j$, the local electronic Green's function $G_j$ becomes a stochastic quantity itself, and one must solve the electron dynamics in the presence of a time-dependent disorder. This is achieved using a non-equilibrium generalization of statistical DMFT \cite{Miranda2011, Janis1992, Dobrosavljevic1993}. For the infinitely coordinated Bethe lattice, this implies that  the DMFT hybridization function on a given site is determined by an average over the Green's function on the opposite sublattice (see App.~\ref{sec:stat_dmft} for further comments on non-equilibrium statistical DMFT). In the simulations, we explicitly treat $128$ representative sites on the $A$ and $B$ sublattices, which give access on the full distribution of local properties. In order to solve the electron dynamics on timescales which are much longer than the intrinsic electron hopping time, we use a quantum Boltzmann equation consistent with DMFT (see App.~\ref{sec:qbe} for a discussion on the quantum Boltzmann equation) \cite{Picano2021}. The simulation of the electron dynamics is done here explicitly because a priori it is not clear whether electrons locally thermalize in the disordered state. For example, the exact solution of the disordered Falikov-Kimball model predicts non-ergodic behavior if the system remains isolated \cite{Eckstein2008}.

\begin{figure}
\centering \includegraphics[width=0.5\textwidth]{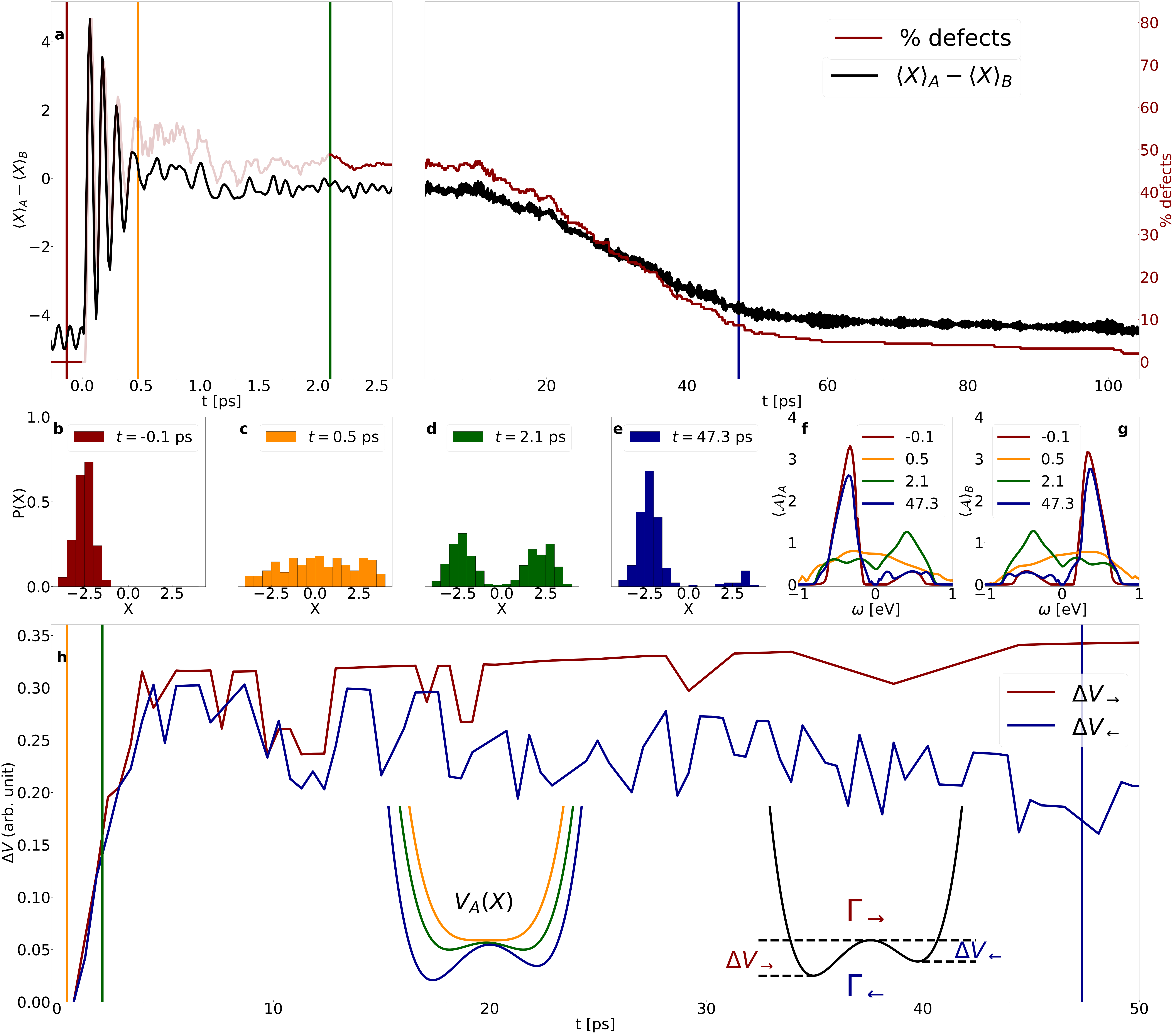}
\caption{ {\bf Characterization of the nonequilibrium dynamics} --- \textbf{a}) Early and long times dynamics (left and right panels, respectively) of $\langle X\rangle_A -\langle X\rangle_B $ and of the defect density $n_{\text{def}}$, normalized to the total number $N_T$ of $A$ and $B$ lattice sites. \textbf{b}) - \textbf{e}) Distribution function of the displacements at different times, as indicated by the vertical lines in panel \textbf{a}). \textbf{f}) - \textbf{g}) Local electronic spectral functions for the $A$ and $B$ sublattices, respectively, at the same times as panels b - e. \textbf{h}) Time evolution of the activation energy from the minority to the majority minimum and back. The left inset shows a sketch of the sublattice potential $V_A(X)$ at different stages of the dynamics as indicated by the vertical lines. The right inset sketches the rates $\Gamma_{\rightarrow}$  ($\Gamma_{\leftarrow}$) for the transition from minority to majority (majority to minority) displacements; $\Delta V_{\rightarrow (\leftarrow)}$ are the barrier heights for the transitions in the two directions.}
\label{fig:disc}
\end{figure}

\section{Numerical simulations} \label{sec:num_sim} 
In the following, we choose representative parameters, with the noninteracting bandwidth $W=1$eV to define the energy scale, $\Omega=0.05$eV (bare phonon period $\tau \sim 80$fs), and $g=0.1$eV (corresponding to a dimensionless coupling constant $\lambda = 4 g^2 /(\Omega W) = 0.8$). All energies are understood in eV unless otherwise stated. We first analyze the equilibrium properties of the model. To prepare an equilibrium state, we initialize the trajectories with $X_j<0$ ($X_j>0$) for $j \in A$ ($j \in B$), and let the system evolve sufficiently long to become stationary on average. Figure \ref{fig:equi}{\bf a} and {\bf b} show the stationary stochastic evolution of the trajectories $X_j$ at all representative sites $j$ at two values of the temperature. At the lower temperature $T = 29$K (Fig.~\ref{fig:equi}{\bf a}), the system is in the ordered phase, and the displacements on the $A$ and $B$ sublattices fluctuate around nonzero values of opposite sign, $\langle X\rangle_A  \approx -2$ and $\langle X\rangle_B \approx 2$. ($\langle \cdot \rangle_\alpha = \langle\cdot \rangle_{j\in \alpha}$ denotes the average over all sites in sublattice $\alpha=A,B$.) At the higher temperature $T = 967$K (Fig.~\ref{fig:equi}{\bf b}), the system is in the disordered phase, and the displacements at both sublattices fluctuate around $X=0$. The dependence of the average $\langle X \rangle_A$ and $\langle X \rangle_B$ indicates a second order phase transition at $T_c \approx 527$K (Fig.~\ref{fig:equi}{\bf c}). The local electronic density of states $\langle \mathcal{A} (\omega) \rangle_\alpha$ at the two sublattices shows a gap in the ordered phase (Fig.~\ref{fig:equi}{\bf d}), which is closed for $T>T_c$ (Fig.~\ref{fig:equi}{\bf e}). Both in the ordered and disordered phase, the fluctuations of the displacement imply that the spectra are substantially broadened with respect to a static mean field simulation with a homogeneous order parameter (green solid lines).

Starting from the insulating solution at $T = 145$K, we now analyze the nonequilibrium dynamics induced by a time-dependent protocol that simulates the photo-excitation of the electrons from the lower to the upper energy band. The transfer of electrons is realized by coupling an electron reservoir with occupied density of states $\mathcal{A}_{\rm dop}^< (\omega)$ at positive energies and unoccupied density of states $\mathcal{A}_{\rm dop}^> (\omega)$ at negative energies (shaded areas in Fig.~\ref{fig:noneq}{\bf a}), for  about $3$fs starting at $t=0$ (see App.~\ref{sec:ex_prot} for details concerning the excitation protocol). At early times, the average distribution function $\langle F (\omega, t) \rangle_A$ assumes a non-equilibrium shape that cannot be fitted by a Fermi-Dirac function. However, it recovers the original form on a relatively short timescale of about $1$ps. (In the figure, we exemplarily show quantities on sublattice $A$; quantities on the other sublattice behave analogously.) The early time electron dynamics can be analyzed in terms of the excitation density, given by the integrated occupation at positive frequencies, $n_A^{\text{ex}} (t) = \int_{0}^{\infty} d \omega \langle F (\omega, t) \mathcal{A} (\omega, t) \rangle_{A}$ (Fig.~\ref{fig:noneq}{\bf b}). Starting from a peak excitation density of few percent, an almost complete decay of $n_A^{\text{ex}}$ due to energy dissipation of the electrons occurs within $\sim 1.3$ps. The average distortion $\langle X\rangle_A - \langle X\rangle_B$ rapidly collapses to zero after the electronic excitation, with an overshoot that represents a strongly damped coherent dynamics  (Fig.~\ref{fig:noneq}{\bf b}). After that, $\langle X\rangle_A - \langle X\rangle_B $ remains close to zero for almost ten times the electronic recovery time, up to $\sim 10$ ps (Fig.~\ref{fig:disc}{\bf a}). For smaller excitation densities, the order parameter is only partially suppressed, and subsequently recovers from that value (see Supplemental Material \cite{Suppl_mat}).

The evolution of $\langle X \rangle_A$ on a longer time scale (Fig.~\ref{fig:noneq}{\bf c}) reveals a very slow relaxation dynamics. If the order is sufficiently suppressed  during the excitation, the final sign of the order parameter becomes random; while the original sign of the order parameter is recovered in the realization shown in Fig 2c, other noise realizations can lead to a reverse (see Supplementary information \cite{Suppl_mat}). The slow dynamics can be linked to the presence of long-living lattice defects, i.e., displacements $X_{j}$ at specific sites $j$ which assume an opposite value with respect to the average displacement on the sublattice of $j$. Indeed, even though $X_{j}$ follows the average $\langle X\rangle_A$ for the majority of sites $j \in A$, a few sites behave differently (see curves $j=0$, $j=16$, and $j=46$ in Fig.~\ref{fig:noneq}{\bf c}). Due to the stochastic nature of the time evolution of the displacements, one can observe trajectories that flip back and forth from positive to negative values, in particular for the early phase of the dynamics. To measure the defect density, we define $n_{\text{def}}$ at a given time by counting the percentage of sites $j$ for which $X_j$ has opposite sign compared to the average $\langle X \rangle_\alpha$ on the given sublattice. Immediately after the photo-excitation, $n_{\text{def}}$ grows from zero (in the original equilibrium state) to a value close to $50 \%$ as $\langle X \rangle_A$ and $\langle X \rangle_B$ drop to zero (Fig.~\ref{fig:disc}{\bf a}). Subsequently, $n_{\text{def}}$ decreases again, but some defects remain even at the latest time of our simulation ($t \sim 100$ps). Given the large separation of time scales between the electronic and the lattice recovery, we expect the inclusion of a small electron-electron interaction not to alter significantly the previous picture. A more general discussion of the role of the different parameters on the dynamics of the system can be found in the Supplemental Material \cite{Suppl_mat}.

To further interpret the data one can look at the full distribution function $P_\alpha (X)$ of the local displacements $X$ on a given sublattice $\alpha=A,B$, which is shown for various times in Fig.~\ref{fig:disc}{\bf b}-{\bf e}. Because $P_A(X)=P_B(-X)$, we show the symmetrized  $P(X) = \big( P_A (X) + P_B (-X) \big)/2$. In the initial equilibrium state, $P(X)$ is peaked around the mean order parameter $\langle X\rangle_A \approx -2$ (Fig.~\ref{fig:disc}{\bf b}). Shortly after the excitation, $P(X)$ first becomes parity-symmetric with a broad distribution around $X = 0$ (Fig.~\ref{fig:disc}{\bf c}). During the relaxation, the distribution then develops into a well-defined bimodal form, which is still parity-invariant, but has reduced weight at $X = 0$ (Fig.~\ref{fig:disc}{\bf d}). This bimodal distribution is the hallmark of the inhomogeneously disordered state. At longer times, the asymmetric equilibrium distribution recovers through a gradual depletion of the minority peak (Fig.~\ref{fig:disc}{\bf e}). 

The evolution of the lattice displacements is indirectly reflected in the spectral function, which can be observed, e.g., in photoemission spectroscopy. The initial reduction of $\langle X\rangle_A$ leads to rapid closing of the energy gap in the original spectral function $\langle \mathcal{A} (\omega, t) \rangle_{\alpha}$, see curve for time  $t=0.5$ps in Fig.~\ref{fig:disc}{\bf f}-{\bf g} for the $A$ and $B$ sublattices, respectively. After the recovery of the nonzero average lattice displacements, the gap is partly restored. However, the presence of the defects leads to incoherent spectral weight in the gap, which remains visible even for the longest simulation time $t=100$ps (Fig.~\ref{fig:noneq}{\bf d}).

\section{Nonequilibrium potential energy barrier} \label{sec:noneq_pot}
The long lifetime of the defects suggests that they are protected by an energy barrier in the effective potential $V_\alpha (X)$ which describes the local dynamics of $X$ on sublattice $\alpha$. For late times, when $P(X)$ shows a bimodal form, we can assume that $V_\alpha (X)$ is represented by a double-well with minima for the majority and minority displacements (see right inset in Fig.~\ref{fig:disc}{\bf h}). One can try to estimate these barriers from the stochastic dynamics, assuming that the rates $\Gamma_{\rightarrow}$  ($\Gamma_{\leftarrow}$) for the transition from minority to majority (majority to minority) are given by the Arrhenius law $\Gamma_{\leftarrow (\rightarrow)} = \Gamma_0 e^{-\Delta V_{\leftarrow (\rightarrow)}/T}$. The barrier heights $\Delta V_{\leftarrow (\rightarrow)}$ for the transitions in the two directions are shown in Fig.~\ref{fig:disc}{\bf h} (see App.~\ref{sec:en_bar} for the extraction of the barriers). The dynamics of the energy barriers is characterized by two stages: Up to a time $t \approx 5$ps before the symmetry breaking sets in, the potential is given by a symmetric double well ($\Delta V_\leftarrow = \Delta V_\rightarrow$), whose barrier increases almost linearly with time. Once the symmetry is broken, the barrier $\Delta V_\leftarrow$ protecting the minority sites remains intact (even if it slowly decreases), but it is now lower than the barrier for the other direction ($\Delta V_\leftarrow < \Delta V_\rightarrow$), supporting the metastable character of the defects.

An important question is the validity of the white noise approximation made in our simulation. At equilibrium, we can compare our results to a DMFT simulation which treats the electron-phonon in leading order perturbation theory but keeps the full retardation effects \cite{Randi2017}. In the high-temperature phase, where the displacements fluctuate around $X=0$, this approach agrees remarkably well with the present simulation regarding the broadening of the equilibrium spectra (see dashed line in Fig.~\ref{fig:equi}{\bf e}).  However, the weak coupling description  cannot reproduce the bimodal phonon distribution by construction and is blind to the inhomogeneous disordering, which sets in in the non-equilibrium state. An exact solution of the Holstein model in DMFT is possible in equilibrium using Quantum Monte Carlo (QMC) techniques \cite{Assaad2007,Werner2007}. Above the ordering temperature, QMC has predicted a highly non-Gaussian and even bimodal distribution $P(X)$ for the Hubbard-Holstein model \cite{Assaad2007}. (Deep in the ordered phase, we expect that the minority displacement in equilibrium has a rather small weight and might be difficult to detect.) While non-equilibrium QMC simulations for the long time dynamics are not possible, our results are clearly in line with these equilibrium findings. Finally, we expect the white noise limit to become systematically better for smaller phonon frequencies. We have performed simulations at a lower $\Omega = 0.005$, corresponding to an oscillation period $\tau \sim 830$fs (keeping the parameter $g^2/\Omega$ fixed, which determines the phase transition in equilibrium). One observes more coherent dynamics at early times (few coherent oscillations of $\langle X \rangle_{A}$ around $X=0$ after the excitation) followed by a fast recovery at later times which indicates no bimodal distribution at intermediate times, and thus a negligible value of the potential energy barrier in this case (see Supplemental Material \cite{Suppl_mat}).

\section{Conclusions} \label{sec:concl} 
In conclusion, we have shown that the dynamics in a simple Holstein model for a Peierls charge density wave transition can lead to ultrafast inhomogeneous disordering after photo-excitation, with a bimodal distribution of the local order parameter. Experimental signatures for a transiently disordered state may be found using scattering techniques \cite{Wall2018}, or, as our model suggests, by means of incoherent spectral weight in the electronic spectra. One might also consider optical experiments which are sensitive to local properties; e.g., if local Raman-active modes are affected by the local order parameter, one could map the order parameter distribution onto a non-trivial distribution of frequencies \cite{Fausti2009} in a stimulated Raman measurement. The stabilization of the disorder is understood in terms of metastable defects, which are local in nature and therefore different from the topologically stabilized defects in the Kibble-Zurek mechanism. The occurrence of inhomogeneous disordering already in the simple  Holstein model suggests this phenomenon should be relevant to photo-induced dynamics in solids more generally. In more complex systems, the inhomogeneous disorder at short times might be a step towards true metastability and glassy states at long times \cite{Gerasimenko2019}. Future theoretical studies will extend the technique developed in this work, i.e., nonequilibrium statistical DMFT with a stochastic lattice evolution, to more realistic descriptions of the coupled electron lattice dynamics, such as a model for VO$_2$ \cite{Grandi2019}.

\begin{acknowledgements}
We thank Simon Wall and Philipp Werner for useful discussions. We acknowledge financial support from the ERC starting grant No.~716648. The numerical calculations have been performed at the RRZE of the University Erlangen-Nuremberg. 
\end{acknowledgements}

A.P. wrote the nonequilibrium DMFT code and performed the numerical simulations with inputs from F.G. and M.E. All the authors contributed to the derivation of the stochastic semiclassical equations for the lattice displacements. F.G. and M.E. conceived the project and wrote the manuscript.

\appendix

\section{Statistical DMFT} \label{sec:stat_dmft}
In DMFT, the Hamiltonian (\ref{hamilt}) is mapped to a set of Anderson-Holstein impurity problems, one for each lattice site $j$ \cite{Georges1996}. These models are defined by the action $S_j = S_j^{loc}+S_j^{hyb}$, where $S_j^{loc}$ describes the coupled electronic and lattice degrees of freedom at the isolated site $j$, and $S_j^{hyb}=\sum_{\sigma} \int_{\mathcal{C}} dt\,dt' c^*_{j, \sigma} (t) \Delta_j (t,t') c_{j, \sigma} (t')$ is the hybridization of the electrons with a self-consistent environment defined through the hybridization function $\Delta_j$. The action is formulated on the Keldysh time contour $\mathcal{C}$
to describe real-time dynamics (see Refs.~\cite{Aoki2014,KamenevBook} for an  introduction to the Keldysh formalism). On the Bethe lattice with coordination number $Z\to\infty$ and nearest neighbor hopping $J_0/\sqrt{Z}$, $\Delta_j$ is given by the average of the local Green's functions $G_m$ at all neighbor sites $m$ of $j$, $\Delta_j (t,t') = |J_0|^2 \sum_{m \in NN(j)} G_m (t,t')/Z$ \cite{Georges1996}. In the present case, we allow all sites to be inequivalent due to a stochastic displacement of the phonons (see below). For  $Z\to\infty$ the sum over nearest neighbors of a site $j$ on the $A$ ($B$) sublattice can be replaced by a statistical average of the respective quantity on the opposite sublattice $B$ ($A$), so that $\Delta_{A(B)} (t,t') = |J_0|^2 \langle G_j(t,t') \rangle_{B(A)}$; $|J_0|=W/4$ is the quarter bandwidth of the noninteracting density of states. In the simulation, we keep $128$ representative impurity models (representative sites) for each sublattice, and evaluate the statistical average accordingly.

\section{Stochastic equation for the lattice distortion} \label{sec:sto_eq}
To solve the impurity model, we separate its action as $S_j = S^{c} + S^{cx}$, where $S^{c} $ contains all purely electronic terms (local contributions and hybridization), and 
\begin{align} 
	S_j^{cx} = & - \int_{\mathcal{C}} dt \ X_j \Big[ \frac{(\partial_t^2 +\Omega^2)}{2 \Omega} X_j + \sqrt{2} g (n_j-1)  \Big]
\end{align}
is the Keldysh action for the displacement $X_j$. The latter describes the uncoupled dynamics of $X_j$ and the coupling to the density $n_j=\sum_\sigma c^*_{j\sigma} c_{j\sigma}$. To find an effective equation of motion for $X_j$ which still takes into account the electronic fluctuations, we closely follow Ref.~\cite{KamenevBook} for the derivation of the Langevin equation for a damped harmonic oscillator: The electrons are integrated out to obtain an action of $X_j$ only, $X_j$  is separated into ``classical'' and ``quantum'' components $X^{\text{cl}}_j$ and $X_j^{\text{q}}$, and  quadratic fluctuations in $X_j^{\text{q}}$ are eliminated in favor of a Gaussian noise through a Hubbard-Stratonovich transformation. The white noise limit is taken because electronic timescales are much faster than the phonon dynamics. In summary, this leads to the following description for the coupled electron-lattice dynamics: (i) The local electronic Green's function $G_j(t,t') = -i \langle T_C c_{j\sigma}(t) c_{j\sigma}^\dagger(t') \rangle_{S_j} $, and the connected electronic density correlation function $\Pi_j(t,t') = i \langle  T_C n (t)  n (t') \rangle^{\text{con}}_{S_j} = i G_j (t, t') G_j (t', t)$ are determined by the impurity action for which $X_j$ is replaced by the time-dependent $X^{\text{cl}}_j(t)$.  (ii) $X^{\text{cl}}_j(t)$ is determined by the equation of motion $\ddot{X}^{\text{cl}}_{j} =f_j$, with the stochastic force Eq.~\eqref{force}. (In the main text, we denote $X_j^{\text{cl}} \equiv X_j$ for simplicity.) The coefficients $\gamma_j$ and $K_j$ in $f_j$ are related to the electronic density correlation function $\Pi_j$ through $\gamma_j (t) = 2 g^2 \Omega \Im \big[ \partial_\omega \Pi_j^R (\omega, t) \big] \vert_{\omega = 0} + \gamma_{\text{HO}}$ and $K_j (t) = g^2 \Omega \Im \big[ \Pi_j^K (\omega, t) \big] \vert_{\omega = 0} + 2T \gamma_{\text{HO}}$; here $\Pi_j^R (\omega, t)$ ($\Pi_j^K (\omega, t)$) are the Wigner-transform of the retarded (Keldysh) component of $\Pi_j$. In the expressions, we have also added a weak extrinsic phonon damping $\gamma_{\text{HO}} \sim \frac{1}{26.33}$ps$^{-1}$ and a consistent noise term $2T \gamma_{\text{HO}}$, which accounts for external dissipation to a bath at the initial temperature $T$. A more detailed derivation of these equations in a more general context can be found in Ref.~\cite{Picano2022_arXiv}, which as a benchmark also shows that the stochastic semiclassical approach can accurately reproduce the phonon distributions in the thermally disordered phase over a wide parameter regime, compared to numerically exact Quantum Monte Carlo simulations.

\section{Quantum Boltzmann equation} \label{sec:qbe}
After the quantum phonon is replaced with a classical stochastic one, one still has to solve the electron impurity model with a time-dependent term $ \propto X^{\text{cl}}_j(t) n_j$. The time evolution of the electronic system, on each lattice site, is provided by a quantum Boltzmann equation (QBE) for the local energy distribution function $F_j ( \omega, t ) = G^<_j (\omega, t)/(2 \pi i\mathcal{A}_j (\omega,t))$ \cite{Picano2021}. The QBE gives an equation $\partial_t F_j ( \omega, t ) = I_{j,\omega} [F]$ for the evolution of the distribution, with scattering integral:
\begin{align} \label{scatt_int}
	I_{j,\omega} [F] = & -i \Gamma^<_j( \omega, t )  - F_j ( \omega, t )  2i \,\text{Im}\, \Gamma^R_j (\omega, t),
\end{align}
where $\Gamma_j(\omega, t)=\Sigma_j(\omega, t)+\Delta_j(\omega, t)$. The self energy $\Sigma_j (\omega, t)$  in particular incorporates  the coupling between the local electronic system and a bosonic bath which acts as a heat reservoir. In time, $\Sigma_j (t, t') = g_{\text{ph}}^2 G_j (t, t') D_{\text{ph}} (t, t')$ where $D_{\text{ph}} (t, t')$ is the propagator for noninteracting bosons with Ohmic density of states $\frac{\omega}{4 \omega_{\text{ph}}^2} \exp(- \omega/ \omega_{\text{ph}})$, $\omega_{\text{ph}} = 0.05$ and $g_{\text{ph}}=0.085$ \cite{Dasari2021,Grandi2021Mott}. As the coupling to the bath is treated in the weak-coupling formalism, a non Ohmic bosonic bath should lead to the same qualitative picture presented in the text \cite{Wilner2015,Peronaci2020}, i.e., a cooling of the photo-excited carriers.

\section{Excitation protocol} \label{sec:ex_prot}
In order to simulate a photo-doping excitation, the system is shortly coupled with a fermionic bath with density of states
\begin{align}
	\mathcal{A}_{\rm dop}(\omega)= \mathcal{A} (\omega-0.625) + \mathcal{A} (\omega+0.625)
\end{align}
consisting of two smooth bands with bandwidth $W_{\rm bath} = 1$ around the energies $\omega_0 = \pm 0.625$. We choose $\mathcal{A} (\omega) = \frac{1}{\pi} \cos^2 (\pi \omega /W_{\text{bath}})$ in the interval $[\omega_0-W_{\rm bath}/2,\omega_0+W_{\rm bath}/2]$. The occupied and unoccupied density of states have spectral shapes given by $\mathcal{A}_{\rm dop}^{<} (\omega)= \mathcal{A}(\omega-\omega_0)$ and $\mathcal{A}_{\rm dop}^{>} (\omega)= \mathcal{A} (\omega+\omega_0)$, respectively (shaded areas in Fig.~\ref{fig:noneq}{\bf a}). This fermionic bath adds a local contribution to the electronic self-energy in Eq.~(\ref{scatt_int}), given by $\Sigma_{\text{dop}}^{>(<)} (t,\omega) =(-)2\pi i V^2(t) \mathcal{A}_{\text{dop}}^{>(<)} (\omega)$, with time-dependent profile $V(t) = V_0 \sin^2 (\pi t/t_0) \theta (t) \theta (t_0-t) $, where $V_0=0.125$ (which is a function of the fluence of the pulse), and $t_0 \sim 2.63$fs (representing the pulse duration). The same excitation protocol has been already applied in \cite{Li2020} and \cite{Picano2021}. Moreover, the fermion bath coupling can be understood as a microscopic model for laser excitation of electrons to/from higher lying bands via dipolar transition matrix elements \cite{Werner2019,grandi2021ultrafast}.

\section{Determination of the energy barriers} \label{sec:en_bar}
To extract the energy barrier  $\Delta V$ in Fig.~\ref{fig:disc}{\bf h} from the rates $\Gamma_{\rightarrow(\leftarrow)}$ and the Arrhenius law, the constant $\Gamma_0$ has been fixed by assuming that the barrier vanishes directly after the excitation, at $t=500$fs (Fig.~\ref{fig:disc}{\bf c}); $T$ is taken to be the final value $T=145$K for simplicity because electronic distributions quickly relax as shown in Fig.~\ref{fig:noneq}{\bf b}. To obtain the rate $\Gamma_{\leftarrow}$ ($\Gamma_{\rightarrow}$) up to a global factor, we measure, over time intervals of  $\Delta t=526$fs, the average number $n_{\text{maj}}$ and $n_{\text{min}}$ of majority and minority trajectories, as well as the number of flips $\Delta n_{\leftarrow}$ and $\Delta n_{\rightarrow}$ in the two directions. A trajectory is considered as flipped when it previously has been at $X<-1.5$ and arrives at $X>1.5$, and vice versa. With this $\Gamma_{\leftarrow} \propto\Delta n_{\leftarrow}/n_{\text{maj}}$ and $\Gamma_{\rightarrow} \propto \Delta n_{\rightarrow}/n_{\text{min}}$.


\begin{thebibliography}{59}
\expandafter\ifx\csname natexlab\endcsname\relax\def\natexlab#1{#1}\fi
\expandafter\ifx\csname bibnamefont\endcsname\relax
  \def\bibnamefont#1{#1}\fi
\expandafter\ifx\csname bibfnamefont\endcsname\relax
  \def\bibfnamefont#1{#1}\fi
\expandafter\ifx\csname citenamefont\endcsname\relax
  \def\citenamefont#1{#1}\fi
\expandafter\ifx\csname url\endcsname\relax
  \def\url#1{\texttt{#1}}\fi
\expandafter\ifx\csname urlprefix\endcsname\relax\def\urlprefix{URL }\fi
\providecommand{\bibinfo}[2]{#2}
\providecommand{\eprint}[2][]{\url{#2}}

\bibitem[{\citenamefont{Basov et~al.}(2017)\citenamefont{Basov, Averitt, and
  Hsieh}}]{Basov2017}
\bibinfo{author}{\bibfnamefont{D.~N.} \bibnamefont{Basov}},
  \bibinfo{author}{\bibfnamefont{R.~D.} \bibnamefont{Averitt}},
  \bibnamefont{and} \bibinfo{author}{\bibfnamefont{D.}~\bibnamefont{Hsieh}},
  \bibinfo{journal}{Nature Materials} \textbf{\bibinfo{volume}{16}},
  \bibinfo{pages}{1077} (\bibinfo{year}{2017}),
  \urlprefix\url{https://doi.org/10.1038/nmat5017}.

\bibitem[{\citenamefont{de~la Torre et~al.}(2021)\citenamefont{de~la Torre,
  Kennes, Claassen, Gerber, McIver, and Sentef}}]{delaTorre2021}
\bibinfo{author}{\bibfnamefont{A.}~\bibnamefont{de~la Torre}},
  \bibinfo{author}{\bibfnamefont{D.~M.} \bibnamefont{Kennes}},
  \bibinfo{author}{\bibfnamefont{M.}~\bibnamefont{Claassen}},
  \bibinfo{author}{\bibfnamefont{S.}~\bibnamefont{Gerber}},
  \bibinfo{author}{\bibfnamefont{J.~W.} \bibnamefont{McIver}},
  \bibnamefont{and} \bibinfo{author}{\bibfnamefont{M.~A.}
  \bibnamefont{Sentef}}, \bibinfo{journal}{Rev. Mod. Phys.}
  \textbf{\bibinfo{volume}{93}}, \bibinfo{pages}{041002}
  (\bibinfo{year}{2021}),
  \urlprefix\url{https://link.aps.org/doi/10.1103/RevModPhys.93.041002}.

\bibitem[{\citenamefont{Yusupov et~al.}(2010)\citenamefont{Yusupov, Mertelj,
  Kabanov, Brazovskii, Kusar, Chu, Fisher, and Mihailovic}}]{Yusupov2010}
\bibinfo{author}{\bibfnamefont{R.}~\bibnamefont{Yusupov}},
  \bibinfo{author}{\bibfnamefont{T.}~\bibnamefont{Mertelj}},
  \bibinfo{author}{\bibfnamefont{V.~V.} \bibnamefont{Kabanov}},
  \bibinfo{author}{\bibfnamefont{S.}~\bibnamefont{Brazovskii}},
  \bibinfo{author}{\bibfnamefont{P.}~\bibnamefont{Kusar}},
  \bibinfo{author}{\bibfnamefont{J.-H.} \bibnamefont{Chu}},
  \bibinfo{author}{\bibfnamefont{I.~R.} \bibnamefont{Fisher}},
  \bibnamefont{and}
  \bibinfo{author}{\bibfnamefont{D.}~\bibnamefont{Mihailovic}},
  \bibinfo{journal}{Nature Physics} \textbf{\bibinfo{volume}{6}},
  \bibinfo{pages}{681} (\bibinfo{year}{2010}),
  \urlprefix\url{https://doi.org/10.1038/nphys1738}.

\bibitem[{\citenamefont{Sch\"afer et~al.}(2010)\citenamefont{Sch\"afer,
  Kabanov, Beyer, Biljakovic, and Demsar}}]{Schaefer2010}
\bibinfo{author}{\bibfnamefont{H.}~\bibnamefont{Sch\"afer}},
  \bibinfo{author}{\bibfnamefont{V.~V.} \bibnamefont{Kabanov}},
  \bibinfo{author}{\bibfnamefont{M.}~\bibnamefont{Beyer}},
  \bibinfo{author}{\bibfnamefont{K.}~\bibnamefont{Biljakovic}},
  \bibnamefont{and} \bibinfo{author}{\bibfnamefont{J.}~\bibnamefont{Demsar}},
  \bibinfo{journal}{Phys. Rev. Lett.} \textbf{\bibinfo{volume}{105}},
  \bibinfo{pages}{066402} (\bibinfo{year}{2010}),
  \urlprefix\url{https://link.aps.org/doi/10.1103/PhysRevLett.105.066402}.

\bibitem[{\citenamefont{Trigo et~al.}(2019)\citenamefont{Trigo, Giraldo-Gallo,
  Kozina, Henighan, Jiang, Liu, Clark, Chollet, Glownia, Zhu
  et~al.}}]{Trigo2019}
\bibinfo{author}{\bibfnamefont{M.}~\bibnamefont{Trigo}},
  \bibinfo{author}{\bibfnamefont{P.}~\bibnamefont{Giraldo-Gallo}},
  \bibinfo{author}{\bibfnamefont{M.~E.} \bibnamefont{Kozina}},
  \bibinfo{author}{\bibfnamefont{T.}~\bibnamefont{Henighan}},
  \bibinfo{author}{\bibfnamefont{M.~P.} \bibnamefont{Jiang}},
  \bibinfo{author}{\bibfnamefont{H.}~\bibnamefont{Liu}},
  \bibinfo{author}{\bibfnamefont{J.~N.} \bibnamefont{Clark}},
  \bibinfo{author}{\bibfnamefont{M.}~\bibnamefont{Chollet}},
  \bibinfo{author}{\bibfnamefont{J.~M.} \bibnamefont{Glownia}},
  \bibinfo{author}{\bibfnamefont{D.}~\bibnamefont{Zhu}}, \bibnamefont{et~al.},
  \bibinfo{journal}{Phys. Rev. B} \textbf{\bibinfo{volume}{99}},
  \bibinfo{pages}{104111} (\bibinfo{year}{2019}),
  \urlprefix\url{https://link.aps.org/doi/10.1103/PhysRevB.99.104111}.

\bibitem[{\citenamefont{Huber et~al.}(2014)\citenamefont{Huber, Mariager,
  Ferrer, Sch\"afer, Johnson, Gr\"ubel, L\"ubcke, Huber, Kubacka, Dornes
  et~al.}}]{Huber2014}
\bibinfo{author}{\bibfnamefont{T.}~\bibnamefont{Huber}},
  \bibinfo{author}{\bibfnamefont{S.~O.} \bibnamefont{Mariager}},
  \bibinfo{author}{\bibfnamefont{A.}~\bibnamefont{Ferrer}},
  \bibinfo{author}{\bibfnamefont{H.}~\bibnamefont{Sch\"afer}},
  \bibinfo{author}{\bibfnamefont{J.~A.} \bibnamefont{Johnson}},
  \bibinfo{author}{\bibfnamefont{S.}~\bibnamefont{Gr\"ubel}},
  \bibinfo{author}{\bibfnamefont{A.}~\bibnamefont{L\"ubcke}},
  \bibinfo{author}{\bibfnamefont{L.}~\bibnamefont{Huber}},
  \bibinfo{author}{\bibfnamefont{T.}~\bibnamefont{Kubacka}},
  \bibinfo{author}{\bibfnamefont{C.}~\bibnamefont{Dornes}},
  \bibnamefont{et~al.}, \bibinfo{journal}{Phys. Rev. Lett.}
  \textbf{\bibinfo{volume}{113}}, \bibinfo{pages}{026401}
  (\bibinfo{year}{2014}),
  \urlprefix\url{https://link.aps.org/doi/10.1103/PhysRevLett.113.026401}.

\bibitem[{\citenamefont{Neugebauer et~al.}(2019)\citenamefont{Neugebauer,
  Huber, Savoini, Abreu, Esposito, Kubli, Rettig, Bothschafter, Gr\"ubel,
  Kubacka et~al.}}]{Neugebauer2019}
\bibinfo{author}{\bibfnamefont{M.~J.} \bibnamefont{Neugebauer}},
  \bibinfo{author}{\bibfnamefont{T.}~\bibnamefont{Huber}},
  \bibinfo{author}{\bibfnamefont{M.}~\bibnamefont{Savoini}},
  \bibinfo{author}{\bibfnamefont{E.}~\bibnamefont{Abreu}},
  \bibinfo{author}{\bibfnamefont{V.}~\bibnamefont{Esposito}},
  \bibinfo{author}{\bibfnamefont{M.}~\bibnamefont{Kubli}},
  \bibinfo{author}{\bibfnamefont{L.}~\bibnamefont{Rettig}},
  \bibinfo{author}{\bibfnamefont{E.}~\bibnamefont{Bothschafter}},
  \bibinfo{author}{\bibfnamefont{S.}~\bibnamefont{Gr\"ubel}},
  \bibinfo{author}{\bibfnamefont{T.}~\bibnamefont{Kubacka}},
  \bibnamefont{et~al.}, \bibinfo{journal}{Phys. Rev. B}
  \textbf{\bibinfo{volume}{99}}, \bibinfo{pages}{220302}
  (\bibinfo{year}{2019}),
  \urlprefix\url{https://link.aps.org/doi/10.1103/PhysRevB.99.220302}.

\bibitem[{\citenamefont{Beaud et~al.}(2014)\citenamefont{Beaud, Caviezel,
  Mariager, Rettig, Ingold, Dornes, Huang, Johnson, Radovic, Huber
  et~al.}}]{Beaud2014}
\bibinfo{author}{\bibfnamefont{P.}~\bibnamefont{Beaud}},
  \bibinfo{author}{\bibfnamefont{A.}~\bibnamefont{Caviezel}},
  \bibinfo{author}{\bibfnamefont{S.~O.} \bibnamefont{Mariager}},
  \bibinfo{author}{\bibfnamefont{L.}~\bibnamefont{Rettig}},
  \bibinfo{author}{\bibfnamefont{G.}~\bibnamefont{Ingold}},
  \bibinfo{author}{\bibfnamefont{C.}~\bibnamefont{Dornes}},
  \bibinfo{author}{\bibfnamefont{S.-W.} \bibnamefont{Huang}},
  \bibinfo{author}{\bibfnamefont{J.~A.} \bibnamefont{Johnson}},
  \bibinfo{author}{\bibfnamefont{M.}~\bibnamefont{Radovic}},
  \bibinfo{author}{\bibfnamefont{T.}~\bibnamefont{Huber}},
  \bibnamefont{et~al.}, \bibinfo{journal}{Nature Materials}
  \textbf{\bibinfo{volume}{13}}, \bibinfo{pages}{923} (\bibinfo{year}{2014}),
  \urlprefix\url{https://doi.org/10.1038/nmat4046}.

\bibitem[{\citenamefont{Zong et~al.}(2019)\citenamefont{Zong, Dolgirev, Kogar,
  Erge\ifmmode~\mbox{\c{c}}\else \c{c}\fi{}en, Yilmaz, Bie, Rohwer, Tung,
  Straquadine, Wang et~al.}}]{Zong2019}
\bibinfo{author}{\bibfnamefont{A.}~\bibnamefont{Zong}},
  \bibinfo{author}{\bibfnamefont{P.~E.} \bibnamefont{Dolgirev}},
  \bibinfo{author}{\bibfnamefont{A.}~\bibnamefont{Kogar}},
  \bibinfo{author}{\bibfnamefont{E.}~\bibnamefont{Erge\ifmmode~\mbox{\c{c}}\else
  \c{c}\fi{}en}}, \bibinfo{author}{\bibfnamefont{M.~B.} \bibnamefont{Yilmaz}},
  \bibinfo{author}{\bibfnamefont{Y.-Q.} \bibnamefont{Bie}},
  \bibinfo{author}{\bibfnamefont{T.}~\bibnamefont{Rohwer}},
  \bibinfo{author}{\bibfnamefont{I.-C.} \bibnamefont{Tung}},
  \bibinfo{author}{\bibfnamefont{J.}~\bibnamefont{Straquadine}},
  \bibinfo{author}{\bibfnamefont{X.}~\bibnamefont{Wang}}, \bibnamefont{et~al.},
  \bibinfo{journal}{Phys. Rev. Lett.} \textbf{\bibinfo{volume}{123}},
  \bibinfo{pages}{097601} (\bibinfo{year}{2019}),
  \urlprefix\url{https://link.aps.org/doi/10.1103/PhysRevLett.123.097601}.

\bibitem[{\citenamefont{Ichikawa et~al.}(2011)\citenamefont{Ichikawa, Nozawa,
  Sato, Tomita, Ichiyanagi, Chollet, Guerin, Dean, Cavalleri, Adachi
  et~al.}}]{Ichikawa2011}
\bibinfo{author}{\bibfnamefont{H.}~\bibnamefont{Ichikawa}},
  \bibinfo{author}{\bibfnamefont{S.}~\bibnamefont{Nozawa}},
  \bibinfo{author}{\bibfnamefont{T.}~\bibnamefont{Sato}},
  \bibinfo{author}{\bibfnamefont{A.}~\bibnamefont{Tomita}},
  \bibinfo{author}{\bibfnamefont{K.}~\bibnamefont{Ichiyanagi}},
  \bibinfo{author}{\bibfnamefont{M.}~\bibnamefont{Chollet}},
  \bibinfo{author}{\bibfnamefont{L.}~\bibnamefont{Guerin}},
  \bibinfo{author}{\bibfnamefont{N.}~\bibnamefont{Dean}},
  \bibinfo{author}{\bibfnamefont{A.}~\bibnamefont{Cavalleri}},
  \bibinfo{author}{\bibfnamefont{S.-i.} \bibnamefont{Adachi}},
  \bibnamefont{et~al.}, \bibinfo{journal}{Nature Materials}
  \textbf{\bibinfo{volume}{10}}, \bibinfo{pages}{101} (\bibinfo{year}{2011}),
  \urlprefix\url{https://doi.org/10.1038/nmat2929}.

\bibitem[{\citenamefont{Stojchevska et~al.}(2014)\citenamefont{Stojchevska,
  Vaskivskyi, Mertelj, Kusar, Svetin, Brazovskii, and
  Mihailovic}}]{Stojchevska2014}
\bibinfo{author}{\bibfnamefont{L.}~\bibnamefont{Stojchevska}},
  \bibinfo{author}{\bibfnamefont{I.}~\bibnamefont{Vaskivskyi}},
  \bibinfo{author}{\bibfnamefont{T.}~\bibnamefont{Mertelj}},
  \bibinfo{author}{\bibfnamefont{P.}~\bibnamefont{Kusar}},
  \bibinfo{author}{\bibfnamefont{D.}~\bibnamefont{Svetin}},
  \bibinfo{author}{\bibfnamefont{S.}~\bibnamefont{Brazovskii}},
  \bibnamefont{and}
  \bibinfo{author}{\bibfnamefont{D.}~\bibnamefont{Mihailovic}},
  \bibinfo{journal}{Science} \textbf{\bibinfo{volume}{344}},
  \bibinfo{pages}{177} (\bibinfo{year}{2014}).

\bibitem[{\citenamefont{Budden et~al.}(2021)\citenamefont{Budden, Gebert,
  Buzzi, Jotzu, Wang, Matsuyama, Meier, Laplace, Pontiroli, Ricc{\`o}
  et~al.}}]{Budden2021}
\bibinfo{author}{\bibfnamefont{M.}~\bibnamefont{Budden}},
  \bibinfo{author}{\bibfnamefont{T.}~\bibnamefont{Gebert}},
  \bibinfo{author}{\bibfnamefont{M.}~\bibnamefont{Buzzi}},
  \bibinfo{author}{\bibfnamefont{G.}~\bibnamefont{Jotzu}},
  \bibinfo{author}{\bibfnamefont{E.}~\bibnamefont{Wang}},
  \bibinfo{author}{\bibfnamefont{T.}~\bibnamefont{Matsuyama}},
  \bibinfo{author}{\bibfnamefont{G.}~\bibnamefont{Meier}},
  \bibinfo{author}{\bibfnamefont{Y.}~\bibnamefont{Laplace}},
  \bibinfo{author}{\bibfnamefont{D.}~\bibnamefont{Pontiroli}},
  \bibinfo{author}{\bibfnamefont{M.}~\bibnamefont{Ricc{\`o}}},
  \bibnamefont{et~al.}, \bibinfo{journal}{Nature Physics}
  \textbf{\bibinfo{volume}{17}}, \bibinfo{pages}{611} (\bibinfo{year}{2021}),
  \urlprefix\url{https://doi.org/10.1038/s41567-020-01148-1}.

\bibitem[{\citenamefont{Maklar et~al.}(2021)\citenamefont{Maklar, Windsor,
  Nicholson, Puppin, Walmsley, Esposito, Porer, Rittmann, Leuenberger, Kubli
  et~al.}}]{Maklar2021}
\bibinfo{author}{\bibfnamefont{J.}~\bibnamefont{Maklar}},
  \bibinfo{author}{\bibfnamefont{Y.~W.} \bibnamefont{Windsor}},
  \bibinfo{author}{\bibfnamefont{C.~W.} \bibnamefont{Nicholson}},
  \bibinfo{author}{\bibfnamefont{M.}~\bibnamefont{Puppin}},
  \bibinfo{author}{\bibfnamefont{P.}~\bibnamefont{Walmsley}},
  \bibinfo{author}{\bibfnamefont{V.}~\bibnamefont{Esposito}},
  \bibinfo{author}{\bibfnamefont{M.}~\bibnamefont{Porer}},
  \bibinfo{author}{\bibfnamefont{J.}~\bibnamefont{Rittmann}},
  \bibinfo{author}{\bibfnamefont{D.}~\bibnamefont{Leuenberger}},
  \bibinfo{author}{\bibfnamefont{M.}~\bibnamefont{Kubli}},
  \bibnamefont{et~al.}, \bibinfo{journal}{Nature Communications}
  \textbf{\bibinfo{volume}{12}}, \bibinfo{pages}{2499} (\bibinfo{year}{2021}),
  \urlprefix\url{https://doi.org/10.1038/s41467-021-22778-w}.

\bibitem[{\citenamefont{Zong et~al.}(2021)\citenamefont{Zong, Dolgirev, Kogar,
  Su, Shen, Straquadine, Wang, Luo, Kozina, Reid et~al.}}]{Zong2021}
\bibinfo{author}{\bibfnamefont{A.}~\bibnamefont{Zong}},
  \bibinfo{author}{\bibfnamefont{P.~E.} \bibnamefont{Dolgirev}},
  \bibinfo{author}{\bibfnamefont{A.}~\bibnamefont{Kogar}},
  \bibinfo{author}{\bibfnamefont{Y.}~\bibnamefont{Su}},
  \bibinfo{author}{\bibfnamefont{X.}~\bibnamefont{Shen}},
  \bibinfo{author}{\bibfnamefont{J.~A.~W.} \bibnamefont{Straquadine}},
  \bibinfo{author}{\bibfnamefont{X.}~\bibnamefont{Wang}},
  \bibinfo{author}{\bibfnamefont{D.}~\bibnamefont{Luo}},
  \bibinfo{author}{\bibfnamefont{M.~E.} \bibnamefont{Kozina}},
  \bibinfo{author}{\bibfnamefont{A.~H.} \bibnamefont{Reid}},
  \bibnamefont{et~al.}, \bibinfo{journal}{Phys. Rev. Lett.}
  \textbf{\bibinfo{volume}{127}}, \bibinfo{pages}{227401}
  (\bibinfo{year}{2021}),
  \urlprefix\url{https://link.aps.org/doi/10.1103/PhysRevLett.127.227401}.

\bibitem[{\citenamefont{Sun and Millis}(2020)}]{Sun2020}
\bibinfo{author}{\bibfnamefont{Z.}~\bibnamefont{Sun}} \bibnamefont{and}
  \bibinfo{author}{\bibfnamefont{A.~J.} \bibnamefont{Millis}},
  \bibinfo{journal}{Phys. Rev. X} \textbf{\bibinfo{volume}{10}},
  \bibinfo{pages}{021028} (\bibinfo{year}{2020}),
  \urlprefix\url{https://link.aps.org/doi/10.1103/PhysRevX.10.021028}.

\bibitem[{\citenamefont{Dolgirev et~al.}(2020)\citenamefont{Dolgirev, Michael,
  Zong, Gedik, and Demler}}]{Dolgirev2020}
\bibinfo{author}{\bibfnamefont{P.~E.} \bibnamefont{Dolgirev}},
  \bibinfo{author}{\bibfnamefont{M.~H.} \bibnamefont{Michael}},
  \bibinfo{author}{\bibfnamefont{A.}~\bibnamefont{Zong}},
  \bibinfo{author}{\bibfnamefont{N.}~\bibnamefont{Gedik}}, \bibnamefont{and}
  \bibinfo{author}{\bibfnamefont{E.}~\bibnamefont{Demler}},
  \bibinfo{journal}{Phys. Rev. B} \textbf{\bibinfo{volume}{101}},
  \bibinfo{pages}{174306} (\bibinfo{year}{2020}),
  \urlprefix\url{https://link.aps.org/doi/10.1103/PhysRevB.101.174306}.

\bibitem[{\citenamefont{Grandi and Eckstein}(2021{\natexlab{a}})}]{Grandi2021}
\bibinfo{author}{\bibfnamefont{F.}~\bibnamefont{Grandi}} \bibnamefont{and}
  \bibinfo{author}{\bibfnamefont{M.}~\bibnamefont{Eckstein}},
  \bibinfo{journal}{Phys. Rev. B} \textbf{\bibinfo{volume}{103}},
  \bibinfo{pages}{245117} (\bibinfo{year}{2021}{\natexlab{a}}),
  \urlprefix\url{https://link.aps.org/doi/10.1103/PhysRevB.103.245117}.

\bibitem[{\citenamefont{Bauer et~al.}(2015)\citenamefont{Bauer, Babadi, and
  Demler}}]{Bauer2015}
\bibinfo{author}{\bibfnamefont{J.}~\bibnamefont{Bauer}},
  \bibinfo{author}{\bibfnamefont{M.}~\bibnamefont{Babadi}}, \bibnamefont{and}
  \bibinfo{author}{\bibfnamefont{E.}~\bibnamefont{Demler}},
  \bibinfo{journal}{Phys. Rev. B} \textbf{\bibinfo{volume}{92}},
  \bibinfo{pages}{024305} (\bibinfo{year}{2015}),
  \urlprefix\url{https://link.aps.org/doi/10.1103/PhysRevB.92.024305}.

\bibitem[{\citenamefont{Lemonik and Mitra}(2017)}]{Lemonik2017}
\bibinfo{author}{\bibfnamefont{Y.}~\bibnamefont{Lemonik}} \bibnamefont{and}
  \bibinfo{author}{\bibfnamefont{A.}~\bibnamefont{Mitra}},
  \bibinfo{journal}{Phys. Rev. B} \textbf{\bibinfo{volume}{96}},
  \bibinfo{pages}{104506} (\bibinfo{year}{2017}),
  \urlprefix\url{https://link.aps.org/doi/10.1103/PhysRevB.96.104506}.

\bibitem[{\citenamefont{Lemonik and Mitra}(2018)}]{Lemonik2018b}
\bibinfo{author}{\bibfnamefont{Y.}~\bibnamefont{Lemonik}} \bibnamefont{and}
  \bibinfo{author}{\bibfnamefont{A.}~\bibnamefont{Mitra}},
  \bibinfo{journal}{Phys. Rev. B} \textbf{\bibinfo{volume}{98}},
  \bibinfo{pages}{214514} (\bibinfo{year}{2018}),
  \urlprefix\url{https://link.aps.org/doi/10.1103/PhysRevB.98.214514}.

\bibitem[{\citenamefont{Stahl and Eckstein}(2021)}]{Stahl2021}
\bibinfo{author}{\bibfnamefont{C.}~\bibnamefont{Stahl}} \bibnamefont{and}
  \bibinfo{author}{\bibfnamefont{M.}~\bibnamefont{Eckstein}},
  \bibinfo{journal}{Phys. Rev. B} \textbf{\bibinfo{volume}{103}},
  \bibinfo{pages}{035116} (\bibinfo{year}{2021}),
  \urlprefix\url{https://link.aps.org/doi/10.1103/PhysRevB.103.035116}.

\bibitem[{\citenamefont{Randi et~al.}(2017)\citenamefont{Randi, Esposito,
  Giusti, Misochko, Parmigiani, Fausti, and Eckstein}}]{Randi2017}
\bibinfo{author}{\bibfnamefont{F.}~\bibnamefont{Randi}},
  \bibinfo{author}{\bibfnamefont{M.}~\bibnamefont{Esposito}},
  \bibinfo{author}{\bibfnamefont{F.}~\bibnamefont{Giusti}},
  \bibinfo{author}{\bibfnamefont{O.}~\bibnamefont{Misochko}},
  \bibinfo{author}{\bibfnamefont{F.}~\bibnamefont{Parmigiani}},
  \bibinfo{author}{\bibfnamefont{D.}~\bibnamefont{Fausti}}, \bibnamefont{and}
  \bibinfo{author}{\bibfnamefont{M.}~\bibnamefont{Eckstein}},
  \bibinfo{journal}{Phys. Rev. Lett.} \textbf{\bibinfo{volume}{119}},
  \bibinfo{pages}{187403} (\bibinfo{year}{2017}),
  \urlprefix\url{https://link.aps.org/doi/10.1103/PhysRevLett.119.187403}.

\bibitem[{\citenamefont{Wall et~al.}(2018)\citenamefont{Wall, Yang, Vidas,
  Chollet, Glownia, Kozina, Katayama, Henighan, Jiang, Miller
  et~al.}}]{Wall2018}
\bibinfo{author}{\bibfnamefont{S.}~\bibnamefont{Wall}},
  \bibinfo{author}{\bibfnamefont{S.}~\bibnamefont{Yang}},
  \bibinfo{author}{\bibfnamefont{L.}~\bibnamefont{Vidas}},
  \bibinfo{author}{\bibfnamefont{M.}~\bibnamefont{Chollet}},
  \bibinfo{author}{\bibfnamefont{J.~M.} \bibnamefont{Glownia}},
  \bibinfo{author}{\bibfnamefont{M.}~\bibnamefont{Kozina}},
  \bibinfo{author}{\bibfnamefont{T.}~\bibnamefont{Katayama}},
  \bibinfo{author}{\bibfnamefont{T.}~\bibnamefont{Henighan}},
  \bibinfo{author}{\bibfnamefont{M.}~\bibnamefont{Jiang}},
  \bibinfo{author}{\bibfnamefont{T.~A.} \bibnamefont{Miller}},
  \bibnamefont{et~al.}, \bibinfo{journal}{Science}
  \textbf{\bibinfo{volume}{362}}, \bibinfo{pages}{572} (\bibinfo{year}{2018}).

\bibitem[{\citenamefont{Perez-Salinas et~al.}(2022)\citenamefont{Perez-Salinas,
  Johnson, Prabhakaran, and Wall}}]{Perez-Salinas2022_NatComm}
\bibinfo{author}{\bibfnamefont{D.}~\bibnamefont{Perez-Salinas}},
  \bibinfo{author}{\bibfnamefont{A.~S.} \bibnamefont{Johnson}},
  \bibinfo{author}{\bibfnamefont{D.}~\bibnamefont{Prabhakaran}},
  \bibnamefont{and} \bibinfo{author}{\bibfnamefont{S.}~\bibnamefont{Wall}},
  \bibinfo{journal}{Nature Communications} \textbf{\bibinfo{volume}{13}},
  \bibinfo{pages}{238} (\bibinfo{year}{2022}),
  \urlprefix\url{https://doi.org/10.1038/s41467-021-27819-y}.

\bibitem[{\citenamefont{Johnson et~al.}(2022)\citenamefont{Johnson,
  Moreno-Menc\'{\i}a, Amuah, Menghini, Locquet, Giannetti, Pastor, and
  Wall}}]{Johnson2022_PRL}
\bibinfo{author}{\bibfnamefont{A.~S.} \bibnamefont{Johnson}},
  \bibinfo{author}{\bibfnamefont{D.}~\bibnamefont{Moreno-Menc\'{\i}a}},
  \bibinfo{author}{\bibfnamefont{E.~B.} \bibnamefont{Amuah}},
  \bibinfo{author}{\bibfnamefont{M.}~\bibnamefont{Menghini}},
  \bibinfo{author}{\bibfnamefont{J.-P.} \bibnamefont{Locquet}},
  \bibinfo{author}{\bibfnamefont{C.}~\bibnamefont{Giannetti}},
  \bibinfo{author}{\bibfnamefont{E.}~\bibnamefont{Pastor}}, \bibnamefont{and}
  \bibinfo{author}{\bibfnamefont{S.~E.} \bibnamefont{Wall}},
  \bibinfo{journal}{Phys. Rev. Lett.} \textbf{\bibinfo{volume}{129}},
  \bibinfo{pages}{255701} (\bibinfo{year}{2022}),
  \urlprefix\url{https://link.aps.org/doi/10.1103/PhysRevLett.129.255701}.

\bibitem[{\citenamefont{Nandkishore and Huse}(2015)}]{Nandkishore2015}
\bibinfo{author}{\bibfnamefont{R.}~\bibnamefont{Nandkishore}} \bibnamefont{and}
  \bibinfo{author}{\bibfnamefont{D.~A.} \bibnamefont{Huse}},
  \bibinfo{journal}{Annual Review of Condensed Matter Physics}
  \textbf{\bibinfo{volume}{6}}, \bibinfo{pages}{15} (\bibinfo{year}{2015}).

\bibitem[{\citenamefont{Abanin et~al.}(2019)\citenamefont{Abanin, Altman,
  Bloch, and Serbyn}}]{Abanin2019}
\bibinfo{author}{\bibfnamefont{D.~A.} \bibnamefont{Abanin}},
  \bibinfo{author}{\bibfnamefont{E.}~\bibnamefont{Altman}},
  \bibinfo{author}{\bibfnamefont{I.}~\bibnamefont{Bloch}}, \bibnamefont{and}
  \bibinfo{author}{\bibfnamefont{M.}~\bibnamefont{Serbyn}},
  \bibinfo{journal}{Rev. Mod. Phys.} \textbf{\bibinfo{volume}{91}},
  \bibinfo{pages}{021001} (\bibinfo{year}{2019}),
  \urlprefix\url{https://link.aps.org/doi/10.1103/RevModPhys.91.021001}.

\bibitem[{\citenamefont{Carleo et~al.}(2012)\citenamefont{Carleo, Becca,
  Schir{\'o}, and Fabrizio}}]{Carleo2012}
\bibinfo{author}{\bibfnamefont{G.}~\bibnamefont{Carleo}},
  \bibinfo{author}{\bibfnamefont{F.}~\bibnamefont{Becca}},
  \bibinfo{author}{\bibfnamefont{M.}~\bibnamefont{Schir{\'o}}},
  \bibnamefont{and} \bibinfo{author}{\bibfnamefont{M.}~\bibnamefont{Fabrizio}},
  \bibinfo{journal}{Scientific Reports} \textbf{\bibinfo{volume}{2}},
  \bibinfo{pages}{243} (\bibinfo{year}{2012}).

\bibitem[{\citenamefont{Smith et~al.}(2017)\citenamefont{Smith, Knolle,
  Moessner, and Kovrizhin}}]{Smith2017a}
\bibinfo{author}{\bibfnamefont{A.}~\bibnamefont{Smith}},
  \bibinfo{author}{\bibfnamefont{J.}~\bibnamefont{Knolle}},
  \bibinfo{author}{\bibfnamefont{R.}~\bibnamefont{Moessner}}, \bibnamefont{and}
  \bibinfo{author}{\bibfnamefont{D.~L.} \bibnamefont{Kovrizhin}},
  \bibinfo{journal}{Phys. Rev. Lett.} \textbf{\bibinfo{volume}{119}},
  \bibinfo{pages}{176601} (\bibinfo{year}{2017}),
  \urlprefix\url{https://link.aps.org/doi/10.1103/PhysRevLett.119.176601}.

\bibitem[{\citenamefont{Yao et~al.}(2016)\citenamefont{Yao, Laumann, Cirac,
  Lukin, and Moore}}]{Yao2016}
\bibinfo{author}{\bibfnamefont{N.~Y.} \bibnamefont{Yao}},
  \bibinfo{author}{\bibfnamefont{C.~R.} \bibnamefont{Laumann}},
  \bibinfo{author}{\bibfnamefont{J.~I.} \bibnamefont{Cirac}},
  \bibinfo{author}{\bibfnamefont{M.~D.} \bibnamefont{Lukin}}, \bibnamefont{and}
  \bibinfo{author}{\bibfnamefont{J.~E.} \bibnamefont{Moore}},
  \bibinfo{journal}{Phys. Rev. Lett.} \textbf{\bibinfo{volume}{117}},
  \bibinfo{pages}{240601} (\bibinfo{year}{2016}),
  \urlprefix\url{https://link.aps.org/doi/10.1103/PhysRevLett.117.240601}.

\bibitem[{\citenamefont{Lan et~al.}(2018)\citenamefont{Lan, van Horssen,
  Powell, and Garrahan}}]{Lan2018}
\bibinfo{author}{\bibfnamefont{Z.}~\bibnamefont{Lan}},
  \bibinfo{author}{\bibfnamefont{M.}~\bibnamefont{van Horssen}},
  \bibinfo{author}{\bibfnamefont{S.}~\bibnamefont{Powell}}, \bibnamefont{and}
  \bibinfo{author}{\bibfnamefont{J.~P.} \bibnamefont{Garrahan}},
  \bibinfo{journal}{Phys. Rev. Lett.} \textbf{\bibinfo{volume}{121}},
  \bibinfo{pages}{040603} (\bibinfo{year}{2018}),
  \urlprefix\url{https://link.aps.org/doi/10.1103/PhysRevLett.121.040603}.

\bibitem[{\citenamefont{van Horssen et~al.}(2015)\citenamefont{van Horssen,
  Levi, and Garrahan}}]{Horssen2015}
\bibinfo{author}{\bibfnamefont{M.}~\bibnamefont{van Horssen}},
  \bibinfo{author}{\bibfnamefont{E.}~\bibnamefont{Levi}}, \bibnamefont{and}
  \bibinfo{author}{\bibfnamefont{J.~P.} \bibnamefont{Garrahan}},
  \bibinfo{journal}{Phys. Rev. B} \textbf{\bibinfo{volume}{92}},
  \bibinfo{pages}{100305} (\bibinfo{year}{2015}),
  \urlprefix\url{https://link.aps.org/doi/10.1103/PhysRevB.92.100305}.

\bibitem[{\citenamefont{Kemper et~al.}(2015)\citenamefont{Kemper, Sentef,
  Moritz, Freericks, and Devereaux}}]{Kemper2015}
\bibinfo{author}{\bibfnamefont{A.~F.} \bibnamefont{Kemper}},
  \bibinfo{author}{\bibfnamefont{M.~A.} \bibnamefont{Sentef}},
  \bibinfo{author}{\bibfnamefont{B.}~\bibnamefont{Moritz}},
  \bibinfo{author}{\bibfnamefont{J.~K.} \bibnamefont{Freericks}},
  \bibnamefont{and} \bibinfo{author}{\bibfnamefont{T.~P.}
  \bibnamefont{Devereaux}}, \bibinfo{journal}{Phys. Rev. B}
  \textbf{\bibinfo{volume}{92}}, \bibinfo{pages}{224517}
  (\bibinfo{year}{2015}),
  \urlprefix\url{https://link.aps.org/doi/10.1103/PhysRevB.92.224517}.

\bibitem[{\citenamefont{Sentef et~al.}(2016)\citenamefont{Sentef, Kemper,
  Georges, and Kollath}}]{Sentef2016}
\bibinfo{author}{\bibfnamefont{M.~A.} \bibnamefont{Sentef}},
  \bibinfo{author}{\bibfnamefont{A.~F.} \bibnamefont{Kemper}},
  \bibinfo{author}{\bibfnamefont{A.}~\bibnamefont{Georges}}, \bibnamefont{and}
  \bibinfo{author}{\bibfnamefont{C.}~\bibnamefont{Kollath}},
  \bibinfo{journal}{Phys. Rev. B} \textbf{\bibinfo{volume}{93}},
  \bibinfo{pages}{144506} (\bibinfo{year}{2016}),
  \urlprefix\url{https://link.aps.org/doi/10.1103/PhysRevB.93.144506}.

\bibitem[{\citenamefont{Werner et~al.}(2012)\citenamefont{Werner, Tsuji, and
  Eckstein}}]{Werner2012}
\bibinfo{author}{\bibfnamefont{P.}~\bibnamefont{Werner}},
  \bibinfo{author}{\bibfnamefont{N.}~\bibnamefont{Tsuji}}, \bibnamefont{and}
  \bibinfo{author}{\bibfnamefont{M.}~\bibnamefont{Eckstein}},
  \bibinfo{journal}{Phys. Rev. B} \textbf{\bibinfo{volume}{86}},
  \bibinfo{pages}{205101} (\bibinfo{year}{2012}),
  \urlprefix\url{https://link.aps.org/doi/10.1103/PhysRevB.86.205101}.

\bibitem[{\citenamefont{Tsuji et~al.}(2013)\citenamefont{Tsuji, Eckstein, and
  Werner}}]{Tsuji2013}
\bibinfo{author}{\bibfnamefont{N.}~\bibnamefont{Tsuji}},
  \bibinfo{author}{\bibfnamefont{M.}~\bibnamefont{Eckstein}}, \bibnamefont{and}
  \bibinfo{author}{\bibfnamefont{P.}~\bibnamefont{Werner}},
  \bibinfo{journal}{Phys. Rev. Lett.} \textbf{\bibinfo{volume}{110}},
  \bibinfo{pages}{136404} (\bibinfo{year}{2013}),
  \urlprefix\url{https://link.aps.org/doi/10.1103/PhysRevLett.110.136404}.

\bibitem[{\citenamefont{Seo et~al.}(2018)\citenamefont{Seo, Tanaka, and
  Ishihara}}]{Seo2018}
\bibinfo{author}{\bibfnamefont{H.}~\bibnamefont{Seo}},
  \bibinfo{author}{\bibfnamefont{Y.}~\bibnamefont{Tanaka}}, \bibnamefont{and}
  \bibinfo{author}{\bibfnamefont{S.}~\bibnamefont{Ishihara}},
  \bibinfo{journal}{Phys. Rev. B} \textbf{\bibinfo{volume}{98}},
  \bibinfo{pages}{235150} (\bibinfo{year}{2018}),
  \urlprefix\url{https://link.aps.org/doi/10.1103/PhysRevB.98.235150}.

\bibitem[{\citenamefont{Georges et~al.}(1996)\citenamefont{Georges, Kotliar,
  Krauth, and Rozenberg}}]{Georges1996}
\bibinfo{author}{\bibfnamefont{A.}~\bibnamefont{Georges}},
  \bibinfo{author}{\bibfnamefont{G.}~\bibnamefont{Kotliar}},
  \bibinfo{author}{\bibfnamefont{W.}~\bibnamefont{Krauth}}, \bibnamefont{and}
  \bibinfo{author}{\bibfnamefont{M.~J.} \bibnamefont{Rozenberg}},
  \bibinfo{journal}{Rev. Mod. Phys.} \textbf{\bibinfo{volume}{68}},
  \bibinfo{pages}{13} (\bibinfo{year}{1996}),
  \urlprefix\url{https://link.aps.org/doi/10.1103/RevModPhys.68.13}.

\bibitem[{\citenamefont{Aoki et~al.}(2014)\citenamefont{Aoki, Tsuji, Eckstein,
  Kollar, Oka, and Werner}}]{Aoki2014}
\bibinfo{author}{\bibfnamefont{H.}~\bibnamefont{Aoki}},
  \bibinfo{author}{\bibfnamefont{N.}~\bibnamefont{Tsuji}},
  \bibinfo{author}{\bibfnamefont{M.}~\bibnamefont{Eckstein}},
  \bibinfo{author}{\bibfnamefont{M.}~\bibnamefont{Kollar}},
  \bibinfo{author}{\bibfnamefont{T.}~\bibnamefont{Oka}}, \bibnamefont{and}
  \bibinfo{author}{\bibfnamefont{P.}~\bibnamefont{Werner}},
  \bibinfo{journal}{Rev. Mod. Phys.} \textbf{\bibinfo{volume}{86}},
  \bibinfo{pages}{779} (\bibinfo{year}{2014}),
  \urlprefix\url{https://link.aps.org/doi/10.1103/RevModPhys.86.779}.

\bibitem[{\citenamefont{Kamenev}(2011)}]{KamenevBook}
\bibinfo{author}{\bibfnamefont{A.}~\bibnamefont{Kamenev}},
  \emph{\bibinfo{title}{{Field Theory of Non-Equilibrium Systems}}}
  (\bibinfo{publisher}{Cambridge University Press}, \bibinfo{year}{2011}).

\bibitem[{\citenamefont{Picano et~al.}(2022)\citenamefont{Picano, Grandi,
  Werner, and Eckstein}}]{Picano2022_arXiv}
\bibinfo{author}{\bibfnamefont{A.}~\bibnamefont{Picano}},
  \bibinfo{author}{\bibfnamefont{F.}~\bibnamefont{Grandi}},
  \bibinfo{author}{\bibfnamefont{P.}~\bibnamefont{Werner}}, \bibnamefont{and}
  \bibinfo{author}{\bibfnamefont{M.}~\bibnamefont{Eckstein}},
  \emph{\bibinfo{title}{{Stochastic semiclassical theory for non-equilibrium
  electron-phonon coupled systems}}} (\bibinfo{year}{2022}),
  \urlprefix\url{https://arxiv.org/abs/2209.00428}.

\bibitem[{\citenamefont{Miranda and Dobrosavljevic}(2011)}]{Miranda2011}
\bibinfo{author}{\bibfnamefont{E.}~\bibnamefont{Miranda}} \bibnamefont{and}
  \bibinfo{author}{\bibfnamefont{V.}~\bibnamefont{Dobrosavljevic}},
  \emph{\bibinfo{title}{{Dynamical mean-field theories of correlation and
  disorder}}} (\bibinfo{year}{2011}), \eprint{1112.6184}.

\bibitem[{\citenamefont{Jani\v{s} and Vollhardt}(1992)}]{Janis1992}
\bibinfo{author}{\bibfnamefont{V.}~\bibnamefont{Jani\v{s}}} \bibnamefont{and}
  \bibinfo{author}{\bibfnamefont{D.}~\bibnamefont{Vollhardt}},
  \bibinfo{journal}{Phys. Rev. B} \textbf{\bibinfo{volume}{46}},
  \bibinfo{pages}{15712} (\bibinfo{year}{1992}),
  \urlprefix\url{https://link.aps.org/doi/10.1103/PhysRevB.46.15712}.

\bibitem[{\citenamefont{Dobrosavljevi\ifmmode~\acute{c}\else \'{c}\fi{} and
  Kotliar}(1993)}]{Dobrosavljevic1993}
\bibinfo{author}{\bibfnamefont{V.}~\bibnamefont{Dobrosavljevi\ifmmode~\acute{c}\else
  \'{c}\fi{}}} \bibnamefont{and}
  \bibinfo{author}{\bibfnamefont{G.}~\bibnamefont{Kotliar}},
  \bibinfo{journal}{Phys. Rev. Lett.} \textbf{\bibinfo{volume}{71}},
  \bibinfo{pages}{3218} (\bibinfo{year}{1993}),
  \urlprefix\url{https://link.aps.org/doi/10.1103/PhysRevLett.71.3218}.

\bibitem[{\citenamefont{Picano et~al.}(2021)\citenamefont{Picano, Li, and
  Eckstein}}]{Picano2021}
\bibinfo{author}{\bibfnamefont{A.}~\bibnamefont{Picano}},
  \bibinfo{author}{\bibfnamefont{J.}~\bibnamefont{Li}}, \bibnamefont{and}
  \bibinfo{author}{\bibfnamefont{M.}~\bibnamefont{Eckstein}},
  \bibinfo{journal}{Phys. Rev. B} \textbf{\bibinfo{volume}{104}},
  \bibinfo{pages}{085108} (\bibinfo{year}{2021}),
  \urlprefix\url{https://link.aps.org/doi/10.1103/PhysRevB.104.085108}.

\bibitem[{\citenamefont{Eckstein and Kollar}(2008)}]{Eckstein2008}
\bibinfo{author}{\bibfnamefont{M.}~\bibnamefont{Eckstein}} \bibnamefont{and}
  \bibinfo{author}{\bibfnamefont{M.}~\bibnamefont{Kollar}},
  \bibinfo{journal}{Phys. Rev. Lett.} \textbf{\bibinfo{volume}{100}},
  \bibinfo{pages}{120404} (\bibinfo{year}{2008}),
  \urlprefix\url{https://link.aps.org/doi/10.1103/PhysRevLett.100.120404}.

\bibitem[{Sup()}]{Suppl_mat}
\bibinfo{note}{See the Supplemental Material
  for the analysis of a different parameter regime with respect to what has
  been analyzed in the main text, for an analysis of the dependence of the
  dynamics by $V_0$, $t_0$ and $g_\text{ph}$ and for the time evolution of the
  stochastic noise and the damping.}

\bibitem[{\citenamefont{Assaad and Lang}(2007)}]{Assaad2007}
\bibinfo{author}{\bibfnamefont{F.~F.} \bibnamefont{Assaad}} \bibnamefont{and}
  \bibinfo{author}{\bibfnamefont{T.~C.} \bibnamefont{Lang}},
  \bibinfo{journal}{Phys. Rev. B} \textbf{\bibinfo{volume}{76}},
  \bibinfo{pages}{035116} (\bibinfo{year}{2007}),
  \urlprefix\url{https://link.aps.org/doi/10.1103/PhysRevB.76.035116}.

\bibitem[{\citenamefont{Werner and Millis}(2007)}]{Werner2007}
\bibinfo{author}{\bibfnamefont{P.}~\bibnamefont{Werner}} \bibnamefont{and}
  \bibinfo{author}{\bibfnamefont{A.~J.} \bibnamefont{Millis}},
  \bibinfo{journal}{Phys. Rev. Lett.} \textbf{\bibinfo{volume}{99}},
  \bibinfo{pages}{146404} (\bibinfo{year}{2007}),
  \urlprefix\url{https://link.aps.org/doi/10.1103/PhysRevLett.99.146404}.

\bibitem[{\citenamefont{Fausti et~al.}(2009)\citenamefont{Fausti, Misochko, and
  van Loosdrecht}}]{Fausti2009}
\bibinfo{author}{\bibfnamefont{D.}~\bibnamefont{Fausti}},
  \bibinfo{author}{\bibfnamefont{O.~V.} \bibnamefont{Misochko}},
  \bibnamefont{and} \bibinfo{author}{\bibfnamefont{P.~H.~M.} \bibnamefont{van
  Loosdrecht}}, \bibinfo{journal}{Phys. Rev. B} \textbf{\bibinfo{volume}{80}},
  \bibinfo{pages}{161207} (\bibinfo{year}{2009}),
  \urlprefix\url{https://link.aps.org/doi/10.1103/PhysRevB.80.161207}.

\bibitem[{\citenamefont{Gerasimenko et~al.}(2019)\citenamefont{Gerasimenko,
  Vaskivskyi, Litskevich, Ravnik, Vodeb, Diego, Kabanov, and
  Mihailovic}}]{Gerasimenko2019}
\bibinfo{author}{\bibfnamefont{Y.~A.} \bibnamefont{Gerasimenko}},
  \bibinfo{author}{\bibfnamefont{I.}~\bibnamefont{Vaskivskyi}},
  \bibinfo{author}{\bibfnamefont{M.}~\bibnamefont{Litskevich}},
  \bibinfo{author}{\bibfnamefont{J.}~\bibnamefont{Ravnik}},
  \bibinfo{author}{\bibfnamefont{J.}~\bibnamefont{Vodeb}},
  \bibinfo{author}{\bibfnamefont{M.}~\bibnamefont{Diego}},
  \bibinfo{author}{\bibfnamefont{V.}~\bibnamefont{Kabanov}}, \bibnamefont{and}
  \bibinfo{author}{\bibfnamefont{D.}~\bibnamefont{Mihailovic}},
  \bibinfo{journal}{Nature Materials} \textbf{\bibinfo{volume}{18}},
  \bibinfo{pages}{1078} (\bibinfo{year}{2019}),
  \urlprefix\url{https://doi.org/10.1038/s41563-019-0423-3}.

\bibitem[{\citenamefont{Grandi et~al.}(2020)\citenamefont{Grandi, Amaricci, and
  Fabrizio}}]{Grandi2019}
\bibinfo{author}{\bibfnamefont{F.}~\bibnamefont{Grandi}},
  \bibinfo{author}{\bibfnamefont{A.}~\bibnamefont{Amaricci}}, \bibnamefont{and}
  \bibinfo{author}{\bibfnamefont{M.}~\bibnamefont{Fabrizio}},
  \bibinfo{journal}{Phys. Rev. Research} \textbf{\bibinfo{volume}{2}},
  \bibinfo{pages}{013298} (\bibinfo{year}{2020}),
  \urlprefix\url{https://link.aps.org/doi/10.1103/PhysRevResearch.2.013298}.

\bibitem[{\citenamefont{Dasari et~al.}(2021)\citenamefont{Dasari, Li, Werner,
  and Eckstein}}]{Dasari2021}
\bibinfo{author}{\bibfnamefont{N.}~\bibnamefont{Dasari}},
  \bibinfo{author}{\bibfnamefont{J.}~\bibnamefont{Li}},
  \bibinfo{author}{\bibfnamefont{P.}~\bibnamefont{Werner}}, \bibnamefont{and}
  \bibinfo{author}{\bibfnamefont{M.}~\bibnamefont{Eckstein}},
  \bibinfo{journal}{Phys. Rev. B} \textbf{\bibinfo{volume}{103}},
  \bibinfo{pages}{L201116} (\bibinfo{year}{2021}),
  \urlprefix\url{https://link.aps.org/doi/10.1103/PhysRevB.103.L201116}.

\bibitem[{\citenamefont{Grandi et~al.}(2021)\citenamefont{Grandi, Li, and
  Eckstein}}]{Grandi2021Mott}
\bibinfo{author}{\bibfnamefont{F.}~\bibnamefont{Grandi}},
  \bibinfo{author}{\bibfnamefont{J.}~\bibnamefont{Li}}, \bibnamefont{and}
  \bibinfo{author}{\bibfnamefont{M.}~\bibnamefont{Eckstein}},
  \bibinfo{journal}{Phys. Rev. B} \textbf{\bibinfo{volume}{103}},
  \bibinfo{pages}{L041110} (\bibinfo{year}{2021}),
  \urlprefix\url{https://link.aps.org/doi/10.1103/PhysRevB.103.L041110}.

\bibitem[{\citenamefont{Wilner et~al.}(2015)\citenamefont{Wilner, Wang, Thoss,
  and Rabani}}]{Wilner2015}
\bibinfo{author}{\bibfnamefont{E.~Y.} \bibnamefont{Wilner}},
  \bibinfo{author}{\bibfnamefont{H.}~\bibnamefont{Wang}},
  \bibinfo{author}{\bibfnamefont{M.}~\bibnamefont{Thoss}}, \bibnamefont{and}
  \bibinfo{author}{\bibfnamefont{E.}~\bibnamefont{Rabani}},
  \bibinfo{journal}{Phys. Rev. B} \textbf{\bibinfo{volume}{92}},
  \bibinfo{pages}{195143} (\bibinfo{year}{2015}),
  \urlprefix\url{https://link.aps.org/doi/10.1103/PhysRevB.92.195143}.

\bibitem[{\citenamefont{Peronaci et~al.}(2020)\citenamefont{Peronaci,
  Parcollet, and Schir\'o}}]{Peronaci2020}
\bibinfo{author}{\bibfnamefont{F.}~\bibnamefont{Peronaci}},
  \bibinfo{author}{\bibfnamefont{O.}~\bibnamefont{Parcollet}},
  \bibnamefont{and} \bibinfo{author}{\bibfnamefont{M.}~\bibnamefont{Schir\'o}},
  \bibinfo{journal}{Phys. Rev. B} \textbf{\bibinfo{volume}{101}},
  \bibinfo{pages}{161101} (\bibinfo{year}{2020}),
  \urlprefix\url{https://link.aps.org/doi/10.1103/PhysRevB.101.161101}.

\bibitem[{\citenamefont{Li et~al.}(2020)\citenamefont{Li, Golez, Mazza, Millis,
  Georges, and Eckstein}}]{Li2020}
\bibinfo{author}{\bibfnamefont{J.}~\bibnamefont{Li}},
  \bibinfo{author}{\bibfnamefont{D.}~\bibnamefont{Golez}},
  \bibinfo{author}{\bibfnamefont{G.}~\bibnamefont{Mazza}},
  \bibinfo{author}{\bibfnamefont{A.~J.} \bibnamefont{Millis}},
  \bibinfo{author}{\bibfnamefont{A.}~\bibnamefont{Georges}}, \bibnamefont{and}
  \bibinfo{author}{\bibfnamefont{M.}~\bibnamefont{Eckstein}},
  \bibinfo{journal}{Phys. Rev. B} \textbf{\bibinfo{volume}{101}},
  \bibinfo{pages}{205140} (\bibinfo{year}{2020}),
  \urlprefix\url{https://link.aps.org/doi/10.1103/PhysRevB.101.205140}.

\bibitem[{\citenamefont{Werner et~al.}(2019)\citenamefont{Werner, Eckstein,
  Müller, and Refael}}]{Werner2019}
\bibinfo{author}{\bibfnamefont{P.}~\bibnamefont{Werner}},
  \bibinfo{author}{\bibfnamefont{M.}~\bibnamefont{Eckstein}},
  \bibinfo{author}{\bibfnamefont{M.}~\bibnamefont{Muller}}, \bibnamefont{and}
  \bibinfo{author}{\bibfnamefont{G.}~\bibnamefont{Refael}},
  \bibinfo{journal}{Nature Communications} \textbf{\bibinfo{volume}{10}}
  (\bibinfo{year}{2019}),
  \urlprefix\url{https://doi.org/10.10382Fs41467-019-13557-9}.

\bibitem[{\citenamefont{Grandi and
  Eckstein}(2021{\natexlab{b}})}]{grandi2021ultrafast}
\bibinfo{author}{\bibfnamefont{F.}~\bibnamefont{Grandi}} \bibnamefont{and}
  \bibinfo{author}{\bibfnamefont{M.}~\bibnamefont{Eckstein}},
  \emph{\bibinfo{title}{{Ultrafast metal-to-insulator switching in a strongly
  correlated system}}} (\bibinfo{year}{2021}{\natexlab{b}}),
  \eprint{2104.03644},
    \urlprefix\url{https://arxiv.org/pdf/2104.03644}.

\end{thebibliography}

\newpage~\newpage~

\appendix
\onecolumngrid 

\section*{Supplementary Information for: 'Inhomogeneous disordering at a photo-induced charge density wave transition'}

\begin{figure*}[h]
\centerline{\includegraphics[width=0.8\textwidth]{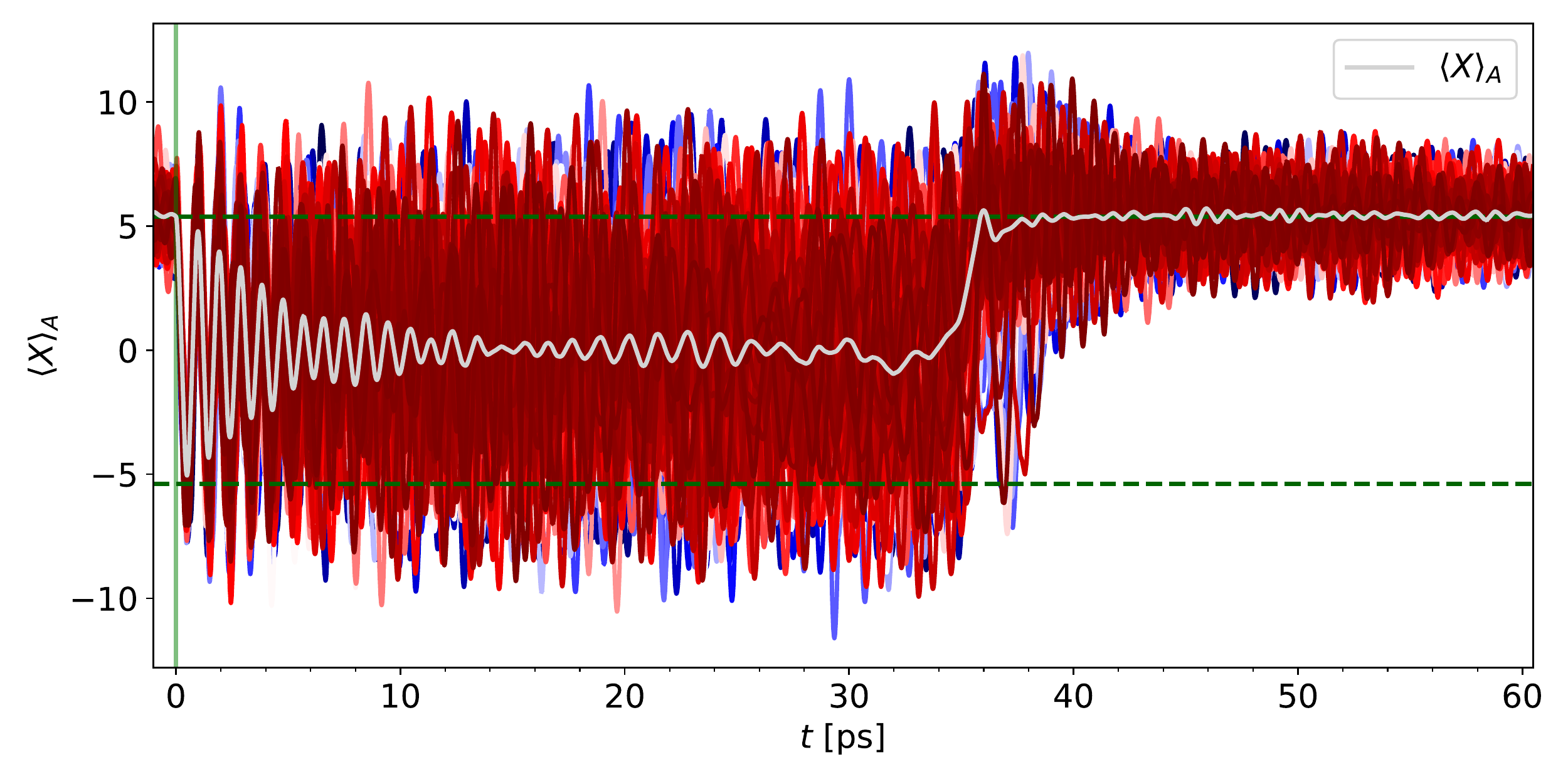}}
\caption{{\bf Nonequilibrium dynamics after photodoping} --- Time evolution of the sublattice $A$ trajectories $X_{j \in A}$ (red to blue lines with different nuances) and of the average $\langle X \rangle_A$.}
\label{fig:nonequisupp}
\end{figure*}

In this Supplemental Information, we present some additional simulations performed in different parameter regimes as compared to the one shown in the main text. Particularly, we 
analyze a different set of parameters for the Hamiltonian of the problem (Sec.~\ref{sec:I}) and different parameters for the laser pulse and the coupling to the phononic bath (Sec.~\ref{sec:II}). Finally, in Sec.~\ref{sec:III}, we show the time evolution of the noise amplitude and the damping parameter for the simulation shown in the main text.

\subsection{Lower phonon frequency}
 \label{sec:I}
In this section, we present some results for the Holstein model [Eq.(1) of the main text] with a smaller phonon frequency than in the main text. We choose a bare frequency $\Omega = 5$meV, and an electron-phonon coupling $g = 26.875$meV. The electronic bandwidth is $W=1$eV, and the ratio $g^2/\Omega$, related to the transition temperature, is the same for this set of parameters and the one considered in the main text. The large value of the oscillation period of the displacements ($\tau \sim 830$fs) obtained in this case leads to a larger difference between the timescales of the electronic and displacement subsystems. The initial temperature is fixed to $T=37$K, where the system lies deep in the insulating state. To excite the system, we use the same protocol as described in the main text (see the Methods). With the new parameters, we observe a dynamics similar to the one presented in the main text. After the electronic photodoping, the average $\langle X \rangle_A$ drops to zero. The initial dynamics $t\lesssim10$ ps shows coherent but damped order parameter oscillations around $X=0$. After $\sim 35$ps, $\langle X \rangle_A$ starts to recover. At $\sim 38$ps, the sublattice average of the displacement has almost completely retrieved its original value ( Fig.~\ref{fig:nonequisupp}). For these values of the parameters, the bimodal distribution observed in the main text disappears, suggesting the absence of the inhomogeneous disordered state in this case.

\subsection{Dependence of the dynamics by $V_0$, $t_0$ and $g_\text{ph}$ } \label{sec:II}
	
In the remaining part of this Supplemental Information, we explore different parameter regimes for the excitation of the system and the strength of the coupling with the dissipative heath bath.
\begin{figure}
	\centerline{\includegraphics[width=0.8\textwidth]{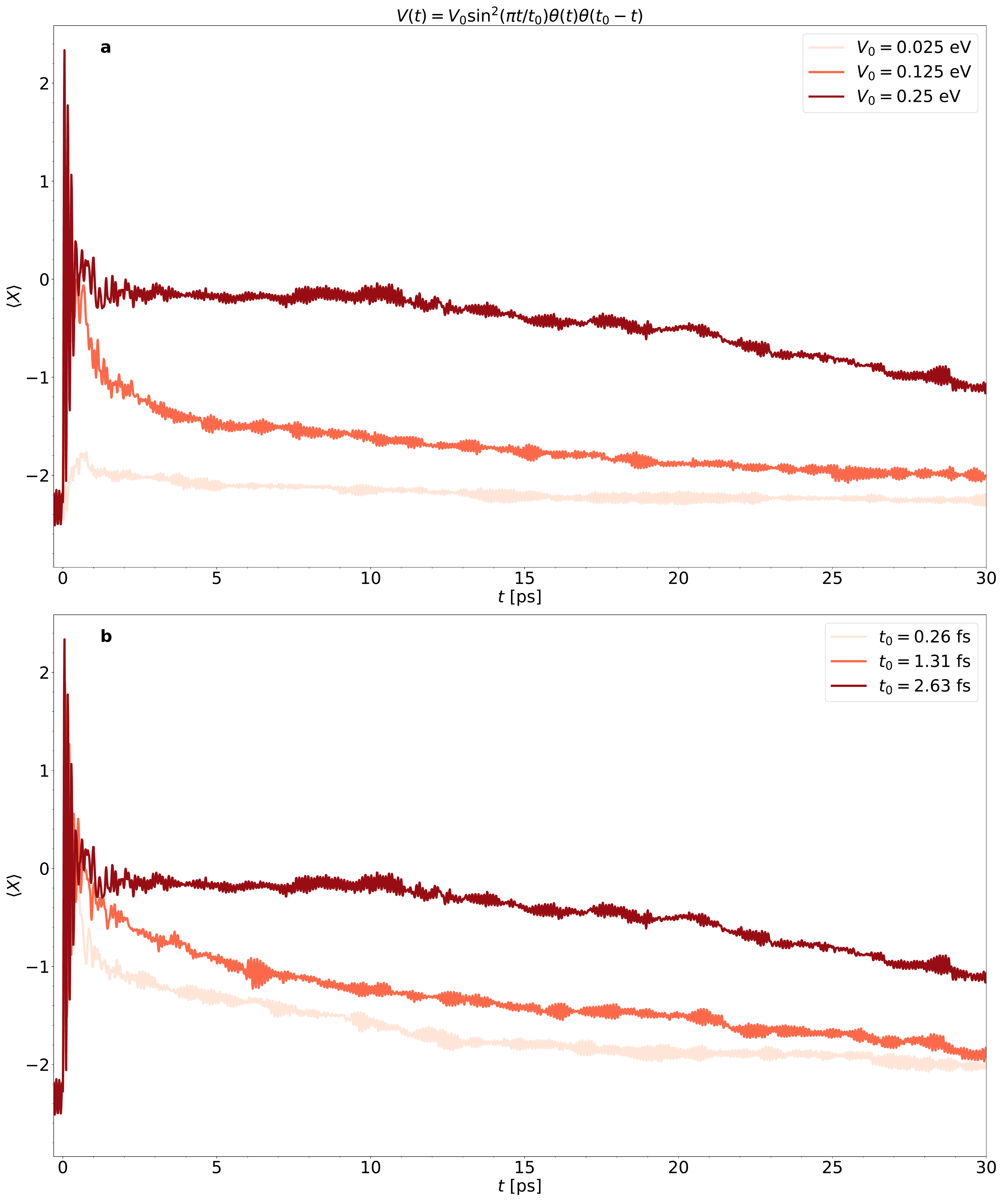}}
	\caption{
	\textbf{a-b}:
	Time evolution of the staggered order parameter $(\langle X \rangle_A$ - $\langle X \rangle_B)/2$ at different amplitudes of the driving field $V_0$ and at different durations $t_0$ of the coupling with the fermionic bath; 
	otherwise, the parameters are identical to the simulation of  Fig.~2-3 in the main text.
	The corresponding values of the excitation density  $n^\text{ex}$, measured at time $\bar t=2.63$fs directly after the excitation, are $n^\text{ex} (\bar t)=0.0046,0.043,0.11$ for the three curves in panel \textbf{a} corresponding, respectively, to $V_0=0.025,0.125,0.25$eV. In panel \textbf{b}, $n^\text{ex} (\bar t)=0.03,0.07,0.11$ correspond, respectively, to durations $t_0=0.26,1.31,2.63$fs. 
	}
	\label{fig:diff_ampl}
\end{figure}
In Fig.~\ref{fig:diff_ampl}a, we show the time evolution of the staggered order parameter 
$(\langle X \rangle_A - \langle X \rangle_B)/2$ 
for several values of the coupling strength $V_0$ to the electron reservoir, where a higher $V_0$ corresponds to a stronger coupling
{(see figure caption for the corresponding values of the excitation density $n^\text{ex}$, measured at time $t=2.63$fs 
directly after the excitation, for the different values of $V_0$)}.
 One can distinguish two regimes for the dynamics of the order parameter. When $V_0 > V_\text{th}$, with $V_\text{th}$ a threshold for the coupling strength, the system shows a complete suppression of the order parameter with coherent oscillations around $\langle X \rangle_A - \langle X \rangle_B \sim 0$ and overshooting dynamics; if instead $V_0 < V_\text{th}$, no overshooting is observed, and the order parameter gets slightly suppressed with a fast recovery of the original amplitude.
Next, we consider the time evolution of the staggered order parameter 
	$(\langle X \rangle_A$ - $\langle X \rangle_B)/2$ 
	for different values $t_0$ of the duration of the excitation. By decreasing $t_0$, we expect a more impulsive (and weaker) perturbation of the system. Results for this analysis are displayed in Fig.~\ref{fig:diff_ampl}b, where a faster recovery of the order parameter by decreasing the duration of the pulse is observed, coherently with the smaller amount of energy that gets transferred to the system.

\begin{figure}
	\centerline{\includegraphics[width=0.8\textwidth]{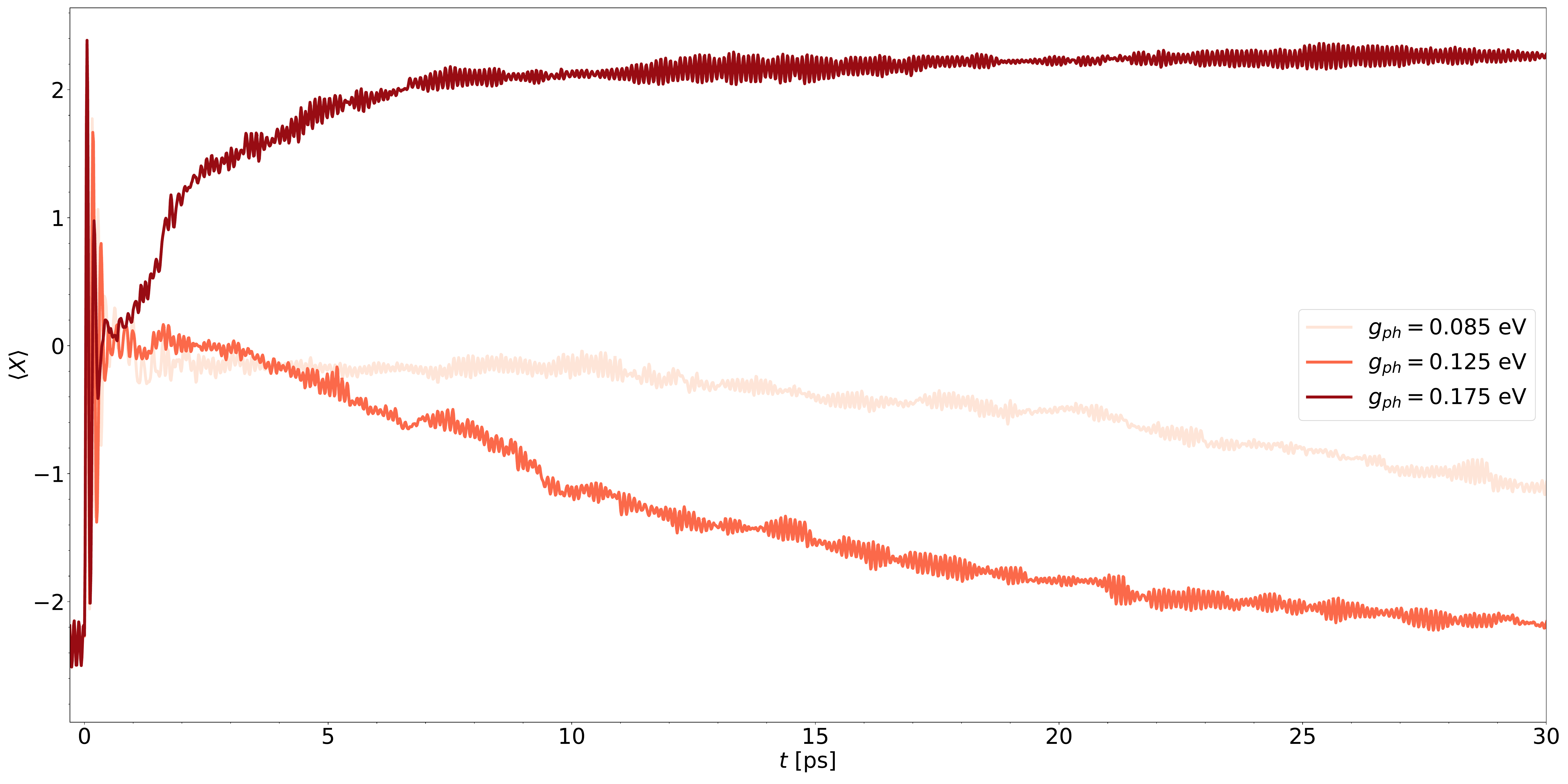}}
	\caption{Time evolution of the staggered order parameter $\langle X \rangle = (\langle X \rangle_A$ - $\langle X \rangle_B)/2$ for different values of the coupling to the phononic bath $g_\text{ph}$; otherwise, the parameters are identical to the simulation of  Fig.~2-3 in the main text. The corresponding values of the excitation density  $n^\text{ex}$, measured at time $\bar t=2.63$fs directly after the excitation, are $n^\text{ex} (\bar t)=0.11,0.054,0.024$ for the three curves in panel corresponding, respectively, to $g_\text{ph}=0.085,0.125,0.175$eV.
	}
	\label{fig:diff_gbath}
\end{figure}

The time evolution of the order parameter for several values of the coupling $g_\text{ph}$ to the phononic (thermal) bath is represented in Fig.~\ref{fig:diff_gbath}. As one would intuitively expect, the larger the coupling, the faster the recovery of the original amplitude of the order parameter.Indeed, a stronger coupling leads to a faster dissipation of the energy injected into the electronic subsystem, which translates in a lower value of the excitation density $n^\text{ex}$ immediately after photodoping (see figure caption for the corresponding values of  $n^\text{ex}$, measured at time $\bar t=2.63$fs).
For $g_{\text ph}=0.175$eV, we notice a sign flip  of $\langle X \rangle_A$ with respect to the original configuration, which occurs randomly. In fact, after the sudden electronic excitation, each of the local lattice deformations $X_j$ is subjected to a different force due to the stochastic term $\xi_j$ in Eq.~(2) of the main text.
When the excitation is strong enough, in the time-domain immediately after the excitation, the lattice distortions oscillate around $\langle X \rangle \approx 0$, suggesting a complete suppression of the ordered state. From this condition, the system spontaneously starts to collapse, as it dissipates the excess energy by coupling with the heath bath, in one of the two degenerate ordered states related by the original $\mathbb{Z}_2$ symmetry of the Hamiltonian (i.e., $\langle X \rangle_A$ starts to collapse towards $+X$ and $\langle X \rangle_B$ towards $-X$, or viceversa). This process is governed by the stochastic contribution to the force that acts on each of the lattice displacements $X_j$, and it can lead to an overall flip of sign of $\langle X \rangle_A$ and $\langle X \rangle_B$ with respect to the original configuration.

\begin{figure}
	\centerline{\includegraphics[width=1.0\textwidth]{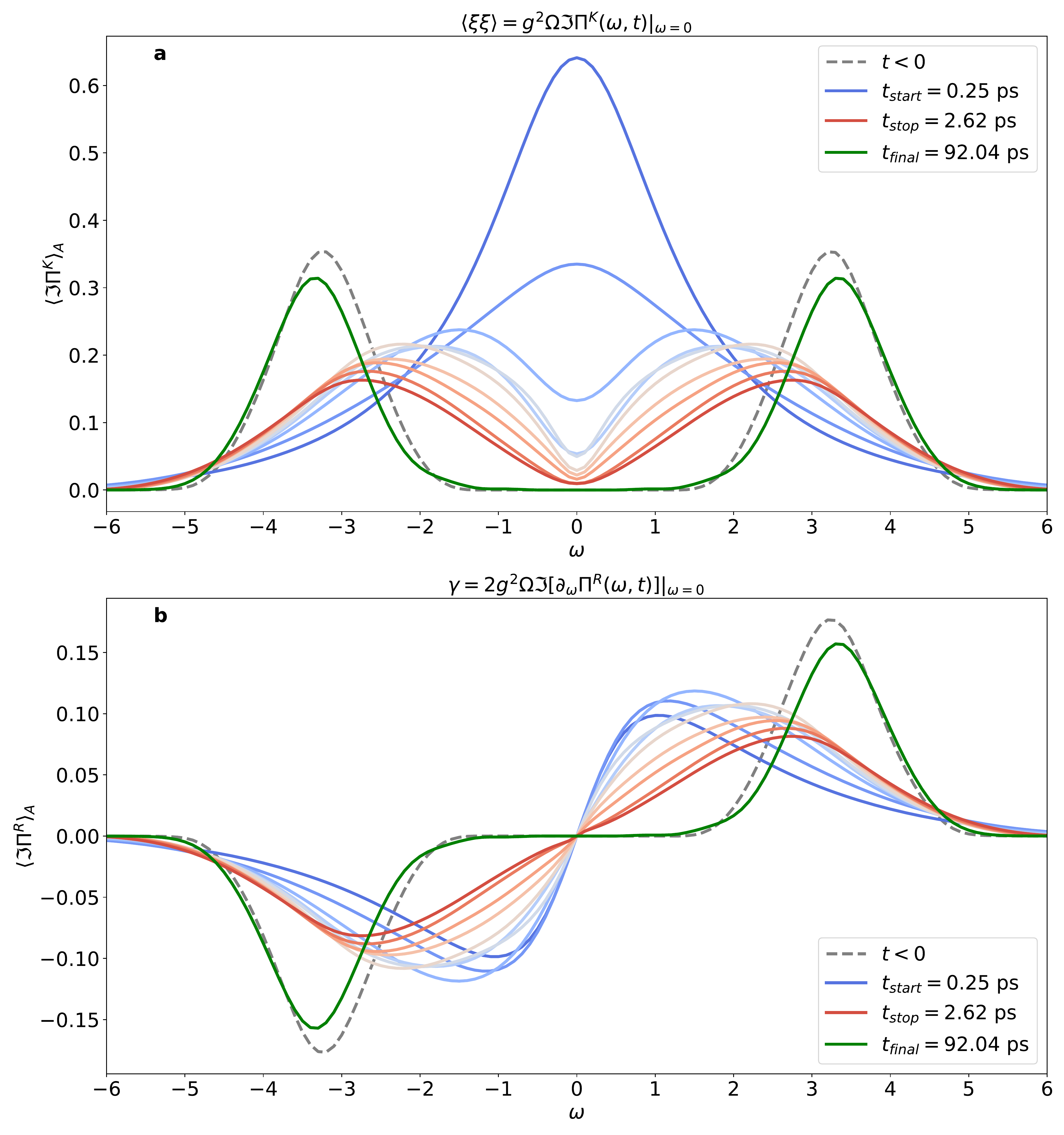}}
	\caption{ 
	\textbf{a-b}: Time evolution of the average, over all the sites, of the connected electronic density correlation function $\Pi$, imaginary Keldysh and Retarded components, respectively, for the same parameters as used in Fig.~2-3 of the main text. $\Im \Pi^K(\omega)$ at frequency $\omega=0$ is proportional to the variance of the Gaussian noise on the phonons; $\Im{[\partial_\omega\Pi^R(\omega)]}|_{\omega=0}$ is proportional to the damping on the phonons. Curves before photoexcitation are shown in grey dashed lines. Solid lines from color blue to red show the evolution of $\Pi$ from $t_{\text{start}}$ to  $t_{\text{stop}}$,  with a time interval $\Delta t \sim 0.26$ps. The curve corresponding to the last step of the time-evolution is shown in green.
	}
	\label{fig:damping_noise}
\end{figure}

\subsection{Time evolution of the stochastic noise and the damping} \label{sec:III}

{
In Fig.~\ref{fig:damping_noise}, we show the time evolution of the 
imaginary part of the retarded and Keldysh components of the connected electronic density correlation function, averaged over all the sites of the $A$ sublattice, 
for the simulation shown in the main text. 
From these quantities, one can see that our method predicts a reduction in the intensity of the stochastic contribution to the force on the phonons and of the average damping, as the system relaxes towards the equilibrium state, as one would expect from the experimental findings. This can be understood qualitatively by considering that the variance of the noise is proportional to the local (i.e., averaged over all the sites of a given sublattice) time-dependent Keldysh component of the Wigner transform of the electronic polarization function computed at $\omega = 0$ (Fig.~\ref{fig:damping_noise}a). For a system close to equilibrium, this is a gapped function with reduced weight at zero frequency. The damping, in turn, is proportional to the local time-dependent derivative with respect to $\omega$ of the retarded component of the Wigner transform of the electronic polarization function, evaluated at $\omega = 0$ (Fig.~\ref{fig:damping_noise}b). For a system that at equilibrium is in the insulating phase, this is a flat function in the region close to zero frequency. 
}

\end{document}